\documentclass[aps,
            showpacs,	
            superscriptaddress,
            nofootinbib]{revtex4-2}
\usepackage[english]{babel}
\usepackage{graphicx}
\usepackage{caption}
\usepackage[labelformat=simple]{subcaption}
\usepackage{natbib}
\usepackage{hyperref}
\usepackage{dcolumn}
\usepackage{bm}
\usepackage{amsmath}
\usepackage{color}
\usepackage{multirow}
\begin{document}

\title{Direct WIMP detection rates for transitions in isomeric nuclei}

\author{M.V Smirnov}
\affiliation{Friedrich-Alexander-Universit\"{a}t Erlangen-N\"{u}rnberg, Erlangen Centre for Astroparticle Physics, Erlangen 91058, Germany}
\affiliation{Department of Physics, Faculty of Mathematics and Natural Sciences, University of Wuppertal, Wuppertal 42119, Germany}
\author{G. Yang}
\email[corresponding author: ]{gyang1@bnl.gov}
\affiliation{Brookhaven National Laboratory, Upton, New York, 11973, USA}
\author{Yu.N. Novikov}
\affiliation{Petersburg Nuclear Physics Institute, Gatchina, St. Petersburg 188300, Russia}
\affiliation{Saint-Petersburg State University, Peterhof, St.Petersburg 198504, Russia}
\author{J.D. Vergados}
\affiliation{Theoretical Physics,University of Ioannina, Ioannina, 45110, Greece}  
\author{D. Bonatsos}
\affiliation{ Institute of Nuclear and Particle
	Physics, National Centre for Scientific
Research “Demokritos”, 15310 Aghia Paraskevi, Attiki, Greece}

\date{\today}

\begin{abstract}
 The direct detection of dark matter constituents, in particular the weakly interacting massive particles (WIMPs), is 
 central to particle physics and cosmology.  
In this paper we study WIMP induced transitions from isomeric nuclear states for two possible isomeric candidates: $\rm^{180}Ta$ and $\rm^{166}Ho$.
The experimental setup, which can measure the possible decay of $\rm^{180}Ta$ induced by WIMPs, was proposed.
The corresponding estimates of the half-life of $\rm^{180}Ta$ are given in the sense that the WIMP-nucleon interaction can be interpreted as ordinary radioactive decay.
\end{abstract}
\pacs{ 95.35.+d, 12.60.Jv 11.30Pb 21.60-n 21.60 Cs 21.60 Ev}


\maketitle

\section{Introduction}
At present there are plenty of evidences of dark matter \textcolor{black}{(DM)} from i) cosmological observations, 
the combined MAXIMA-1~\cite{MAXIMA-1}, BOOMERANG ~\cite{BOOMERANG},
DASI~\cite{DASI}, COBE/DMR Cosmic Microwave Background (CMB)
observations~\cite{COBE, SPERGEL},  
as well as the recent WMAP ~\cite{WMAP06} and Planck~\cite{PlanckCP13} data 
and ii)  the observed rotational curves in the galactic halos, see e.g. the review~\cite{UK01}. It is, however, essential to directly
detect such matter in order to unravel the nature of its constituents.

At the moment, there are many candidates, so-called Weakly Interacting Massive Particles (WIMPs), e.g. the LSP (Lightest Super-symmetric Particle)~\cite{ref2a,ref2b,ref2c,ref2,ELLROSZ,Gomez,ELLFOR}, technibaryon~\cite{Nussinov92,GKS06}, mirror matter~\cite{FLV72,Foot11}, Kaluza-Klein models with universal extra dimensions~\cite{ST02a,OikVerMou} etc. Meanwhile, proposals such as DM as dark photons, axion-like particles and light scalar bosons were studied~\cite{Smirnov:2021wgi}. 
These models predict an interaction of DM with ordinary matter via the exchange of a scalar particle, which leads to a spin independent interaction (SI) or vector boson interaction, which leads to a spin dependent (SD) nucleon cross section.    
Additional theoretical tools are the structure of the nucleus, see e.g.~\cite{JDV06a,Dree00,Dree,Chen}, and the nuclear matrix elements (NMEs)~\cite{Ress,DIVA00,JDV03,JDV04,VF07}. 

In this paper, we will focus on the spin dependent WIMP nucleus interaction. 
This cross section can be sizable in a variety of models, including the lightest super-symmetric particle (LSP)~\cite{CHATTO,CCN03,JDV03, WELLS}, in the co-annihilation region~\cite{Cannoni11}, where the ratio of the SD  to to the SI nucleon cross section, depending on $\tan{\beta}$ and the WIMP mass  can be  large, e.g.  $10^3$ in the WIMP mass range 200-500 GeV.

Furthermore more recent calculations in the super-symmetric $SO(10)$ model~\cite{Gogoladze13}, also in the co-annihilation region, predict ratios of the order of $2\cdot 10^{3}$  for a WIMP mass of about 850 GeV. 
Models of exotic WIMPs, like Kaluza-Klein models~\cite{ST02a,OikVerMou} and Majorana particles with  spin  3/2~\cite{SavVer13}, also can lead to large nucleon spin induced cross sections, which satisfy the relic abundance constraint.  
This interaction is very important because it can lead to inelastic WIMP-nucleus scattering with a non 
a prospect proposed some time ago~\cite{GOODWIT} and considered in some detail by Ejiri and collaborators~\cite{EFO93}.  
Indeed for a Maxwell-Boltzmann (M-B) velocity distribution the average kinetic energy of the WIMP is:
 \begin{equation} 
 \langle T\rangle \approx50~{\rm keV}
\frac{m_{\chi}}{100~{\rm GeV}}
 \label{kinen}
 \end{equation}
 So, for sufficiently heavy WIMPs, the available energy via the high velocity tail of the M-B distribution may be adequate~\cite{EFL88} to allow  scattering  to low lying excited states of certain targets, e.g. of 
 $57.7~$keV for the $7/2^{+}$ excited state of $^{127}$I, the 39.6 keV for the first excited $3/2^{+}$ of $^{129}$Xe, 35.48 keV for the first excited $3/2^+$ state of $^{125}$Te  and 9.4 keV for the first excited $7/2^{+}$ state of $^{83}$Kr.
 In fact calculations of the event rates for the inelastic WIMP-nucleus transitions involving the above systems have been performed~\cite{JDVES13,JDVAPSKS15}.
\textcolor{black}{However, these levels live for a very short time (much less than 1~$\mu s$), and are unsuitable for the expected long-term exposures to search for WIMP-nucleus interactions. 
At the same time, there is a set of long-lived nuclear meta-stable states that can be artificially directly produced in reactors and accelerators and prepared in the form of samples for their long exposure under
low-background experimental conditions.
The possible candidates are listed in Tab.~\ref{tab:1}}

\begin{table}[h!]
\centering
\caption{Long-lived isomeric states with high spin differences that can be effectively produced and used for the DM search.
}
\label{tab:1}
\begin{tabular}{ c|c|c|c|c  }
 \hline
 Isomer & Half-life (year) & Energy (keV) &Spin sequence from I.S. to G.S. & Decay B.R.\\
 \hline
 102mRh   & 3.74    & 140.73 (9)& $6^+$ $\rightarrow$ $2^-$ & $\beta$ $\approx$ 100\%\\
 108mAg&   438      & 109.406 (7) (9)& $6^+$ $\rightarrow$ $1^+$ & Isomer tran. 9\%\\
 110mAg & 0.68       & 117.59 (5)& $6^+$ $\rightarrow$ $2^-$ $\rightarrow$ $1^+$ & Isomer tran. 1.3\%\\
 166mHo    & 1130   & 5.965 (12)& $7^-$ $\rightarrow$ $0^-$ & $\beta$ $\approx$ 100\%\\
 178mHf&  31        & 2446.09 (8)& $16^+$ $\rightarrow$ $8^-$ $\rightarrow$ $0^+$ & Isomer tran. $\approx$ 100\%\\
 180mTa& $>7\cdot10^{15}$ & 76.79 (55)& $9^-$ $\rightarrow$ $1^+$ & \\
 186mRe& $2\cdot10^5$   & 148.2 (5)& $8^+$ $\rightarrow$ $1^-$ & Isomer tran. $>$ 90\%\\
192mIr& 241    & 168.14 (12)& $11^-$ $\rightarrow$ $4^+$ & Isomer tran. $\approx$ 100\%\\
210mBi& $3\cdot10^6$   & 271.31 (11)& $9^-$ $\rightarrow$ $1^-$ & $\alpha$ $\approx$ 100\%\\
242mAm& 141   & 48.60 (5)& $5^-$ $\rightarrow$ $1^-$ & Isomer tran. $\approx$ 100\%\\
 \hline
\end{tabular}
\end{table}
 
Interest in the inelastic WIMP-nucleus scattering has recently been revived by a new proposal to search for the de-excitation of meta-stable nuclear isomers~\cite{PRR20} after such collisions. 
The longevity of these isomers is related to a strong suppression of $\gamma$ and $\beta$-transitions, typically inhibited by a large difference in the angular momentum for the nuclear transition. 
Collisional de-excitation by DM is possible
since heavy DM particles can have a momentum exchange with the nucleus comparable to the
inverse nuclear size, hence lifting tremendous angular momentum suppression of the nuclear transition.

\textcolor{black}{In this work we consider very long-lived isomeric states that can be directly and efficiently produced, then (if necessary) chemically separated and prepared for long-term exposures.
The lifetimes of these isomeric states must be preferably longer than the time of their many-years exposures to detect interactions with DM. 
The baseline of this research is placed on a thorough analysis of the possibility of observing the effect of the interaction of DM particles with isomeric states in $\rm^{166}Ho$ and $\rm^{180}Ta$, of which the first can be obtained in a reactor, and the second is completely quasi-stable and can be separated from a mixture of tantalum isotopes in nature. 
To search for a signal from DM, the cryogenic microcalorimetry method can be used, which has shown its many times better efficiency and accuracy compared to the previously used semiconductor spectroscopy approach.}

\textcolor{black}{
Given the requirement from existing work~\cite{PRR20}, this paper provides detailed calculations for the dedicated nuclear theory evaluation of WIMP induced transition. Meanwhile a new type of detection technique is proposed to perform the experiment. The article is arranged as below. In section II and III, the basic kinematics and cross section formula are laid out. In section IV, the general nuclear structure consideration is introduced. In section V and VI, the detailed cross section calculations for two different targets are shown and in section VII and VIII, an experimental proposal with sensitivity are given, followed by a discussion section in IX.
}
 
 \section{Kinematics}

Evaluation of the differential rate for a WIMP induced  transition $A^i_{iso}(E_x)$  for an isomeric nuclear state at excitation energy $E_x$ to another one  $A^f_{iso}(E'_x)$ (or to the ground state) proceeds in a fashion similar to that of the standard inelastic WIMP induced transition, except in the consideration of kinematics. 
We will make a judicious choice of the final nuclear state that can decay in a standard way to the ground state or to another lower excited state:
\begin{equation}
A^i_{iso}(E_x)+\chi \rightarrow A^f_{iso}(E'_x)+\chi,
\end{equation}
with $\chi$ the DM particle (WIMP). 
Assuming that all particles involved are non relativistic we get:
\begin{equation}
\frac{{\bf p}^2_{\chi}}{2 m_{\chi}}+E_x=\frac{{\bf p}'^2_{\chi}}{2 m_{\chi}}+E'_x+\frac{{\bf q}^2}{2 m_{A}},
\end{equation}
where ${\bf q}$ is the momentum transfer to the nucleus ${\bf q}={\bf p}_{\chi}-{\bf p}'_{\chi}$ and $m_A$ is the isomer mass. So the above equation becomes
\begin{equation}
 \frac{-{\bf q}^2}{2 \mu_r}+\upsilon \xi q -\Delta =0, \quad \Delta=E_x-E'_x \Leftrightarrow -\frac{m_A}{\mu_r}E_R +\upsilon \xi \sqrt{2 m_A E_R}+\Delta=0,
\end{equation}
 where $\Delta>0$, $\xi$ is the cosine of the angle between the incident WIMP and the recoiling nucleus, $\upsilon$ the oncoming WIMP velocity,  $\mu_r = \frac{m_A m_\chi}{m_A + m_\chi}$ the reduced mass of the WIMP-nucleus system and $E_R$ the nuclear recoil energy.
 From the above expression it is immediately apparent that 
\begin{equation}
  \xi=\frac{-\Delta +\frac{m_A}{\mu_r}E_R}{\upsilon \sqrt{2 m_A E_R}}  
\end{equation}
Thus we find the next condition
$$
-1 \le \frac{-\Delta +\frac{m_A}{\mu_r}E_R}{\upsilon \sqrt{2 m_A E_R}}\le 1
$$
This means that for $-\Delta +\frac{m_A}{\mu_r}E_R>0$ the allowed range of velocities is
\begin{equation}
\upsilon_2\le \upsilon\le \upsilon_{\rm esc},\,\upsilon_2=\frac{-\Delta +\frac{m_A}{\mu_r}E_R}{ \sqrt{2 m_A E_R}},
\label{Eq:maxq1}
\end{equation}
where $\upsilon_{\rm esc}$ is the escape velocity, the maximum allowed velocity. 
For $-\Delta +m_A / \mu_r \cdot E_R<0$ the allowed region is
\begin{equation}
\upsilon_1\le \upsilon\le \upsilon_{\rm esc},\,\upsilon_1=\frac{\Delta -\frac{m_A}{\mu_r}E_R}{ \sqrt{2 m_A E_R}}
\end{equation}
At this point we should mention that in the standard inelastic scattering the region  
$$\upsilon \le\frac{E_x+\frac{m_A}{\mu_r}E_R}{ \sqrt{2 m_A E_R}} $$ 
is not available. Note the difference in sign between the previous equation and Eq.\eqref{Eq:maxq1}. As a result  $\upsilon_{min}$ increases with $E_x$, which explains   the suppression of the expected rates in the standard process as $E_x$ increases.

Based on Appendix~\ref{app:rec_energy}, in the special case of WIMP-nucleon scattering the maximum recoil energy is given by 
\begin{equation}
\left(E_R\right)_{\rm max}=2 m_N\upsilon_{\rm esc}^2\frac{1}{(1+x)^2}=2 m_N\upsilon_{0}^2y_{\rm esc}^2\frac{1}{(1+x)^2},
\end{equation}
where $x = m_N / m_\chi$.
Using $\upsilon_0 \approx  0.7\cdot  10^{-3}$ (in natural units) and $y_{\rm esc}=2.84$ we obtain
\begin{equation}
\left(E_R\right)_{\rm max}=8.0\times 10^{-6}m_N  \frac{\rm GeV}{(1+x)^2}\approx8.0 \frac{\rm keV}{(1+x)^2}
\label{ERMax}
\end{equation}

\section{Expressions for the cross section}

The differential cross section is given by 
\begin{equation}
d\sigma=\frac{1}{\upsilon}\frac{1}{(2 \pi)^2}d^3 {\bf q}\delta\left (\frac{q^2}{2 \mu_r}-q \upsilon \xi-\Delta\right )\left (\frac{G_F}{\sqrt{2}}\right )^2|ME(q^2)|^2
\end{equation}
where $|ME(q)|^2$ is the NME of the  WIMP-nucleon interaction in dimensionless units and $G_F$ the standard weak interaction strength.
 Integrating over $\xi$ by making use of the $\delta$ function we get
\begin{equation}
d\sigma=\frac{1}{\upsilon}\frac{1}{(2 \pi)^2}q^2 d q \frac{2 \pi}{q \upsilon}\left (\frac{G_F}{\sqrt{2}}\right )^2|ME(q^2)|^2
\label{Eq.DifCros}
\end{equation}

New physics is contained in elementary nucleon interaction, so we prefer to parameterize in terms of the elementary nucleon cross section.
In the case of the nucleon Eq.\eqref{Eq.DifCros}  becomes
\begin{equation}
d\sigma=\frac{1}{\upsilon^2}\frac{1}{2 \pi}q d q \left (\frac{G_F}{\sqrt{2}}\right )^2|ME_N|^2=\frac{1}{\upsilon^2}\frac{1}{2 \pi} m_N dE_R \left (\frac{G_F}{\sqrt{2}}\right )^2|ME_N|^2
\label{Eq.DifCrosN}
\end{equation}
Folding the last equation with the velocity distribution and integrating over the allowed recoil energies (see the Appendix~\ref{app:NWimpRate}), we obtain
\begin{equation}
\sigma_N=4.0 \frac{1}{2 \pi} m^2_N  \left (\frac{G_F}{\sqrt{2}}\right )^2  \left (f_V^2+3 f_A^2\right )
\label{Eq.TotCrosN2}
\end{equation}

\section{Nuclear structure}

The microscopic structure of atomic nuclei is described in terms of the spherical shell model~\cite{Mayer,Jensen,MJ}, introduced in 1949 in order to explain the magic numbers 2, 8, 20, 28, 50, 82, 126, \dots, at which nuclei present particularly stable configurations. 
The shell model is obtained from the three-dimensional isotropic harmonic oscillator, to which the spin-orbit interaction is added. 
It offers a satisfactory description of nuclei with few valence protons and valence neutrons outside   closed shells, corresponding to the magic numbers, but it fails to explain the experimentally observed large nuclear quadrupole moments away from closed shells, where it has been suggested~\cite{Rainwater} in 1950 that spheroidal instead of spherical shapes lead to greater stability. Along this line, the collective model of Bohr and Mottelson~\cite{Bohr,BM} was introduced in 1952, in which departure from the spherical shape and from axial symmetry are described by the collective variables $\beta$ and $\gamma$, respectively. 
Furthermore in 1955 the Nilsson model~\cite{Nilsson,RN,NR} was introduced, in which a cylindrical harmonic oscillator is used instead of a spherical one, characterized by a deformation $\epsilon$, reflecting the departure of the cylindrical shape from spherical one. 
The single particle orbitals in the Nilsson model are labeled by $\Omega[N n_z \Lambda]$, where $N$ is the total number of the oscillator quanta, $n_z$ is the number of quanta along the 
$z$-axis of cylindrical symmetry, while $\Lambda$ ($\Omega$) is the projection of the orbital (total) angular momentum on the $z$-axis. 

In what follows it will be of interest to consider the expansions of the Nilsson orbitals in the spherical shell model basis $|N l j \Omega \rangle $, where $N$ is the principal quantum number, $l$ ($j$) is the orbital (total) angular momentum, and $\Omega$ is the projection of the total angular momentum on the $z$-axis. 
The necessary expansions have been obtained as described in~\cite{EPJPSM}  and are shown in Tab.~\ref{tab:3} and \ref{tab:4} for three different values of the deformation $\epsilon$.

An important remark on the expansions shown in Table~\ref{tab:3}  is in order. 
One can see that there is a basic difference between intruder orbitals (orbitals pushed within the spherical shell model by the spin-orbit interaction to the oscillator shell below) and normal parity orbitals (orbitals remaining in their own oscillator shell). 
Intruder orbitals remain concentrated on one spherical shell model vector at all deformations, while normal parity orbitals are concentrated on one spherical shell model orbital at small deformation, but will in general be distributed onto several spherical shell model orbitals at large deformations. 
The implications of this difference will become clear below. 
It should be mentioned that the ``purity'' observed in the case of the intruder orbitals is due to the fact that they do not mix with their normal parity neighbors, while the normal parity orbitals, of which there are more than one, do mix among themselves. 

\subsection{The nucleus $^{166}$Ho}
\label{sec:nucleusHo}

The even-even core of $\rm ^{166}_{~67}Ho_{99}$ is $\rm ^{164}_{~66}Dy_{98}$, for which the experimental value of the collective deformation variable $\beta$ is 0.3486~\cite{Pritychenko}, thus the Nilsson deformation 
$\epsilon=0.95 \beta$~\cite{NR} is 0.3312~.  

Covariant density functional theory calculations using the DDME2 functional indicate that the first neutron orbitals lying above the Fermi surface of the core nucleus $\rm^{164}_{~66}Dy_{98}$ are 1/2[521] (lower) and 7/2[633] (higher), while the first proton orbitals lying above the Fermi surface of the core nucleus $\rm^{164}_{~66}Dy_{98}$ are 7/2[523] (lower) and 7/2[404] (higher), with 3/2[411] lying at the Fermi surface.

In 1978 it has been argued\footnote{N. Barron, Ph.D thesis, Louisiana State U. (1978).} that the $0^-$ ground state of  $\rm^{166}_{~67}Ho_{99}$ should arise from the coupling of the 7/2[633] neutron to the 7/2[523] proton. The $7^-$ isomer state can also arise from these orbitals. 

Let us now consider the formation of the above mentioned states under the light of the expansions of the Nilsson orbitals in terms of spherical shell model 
orbitals, shown in Tabs.~\ref{tab:3} and \ref{tab:4}.
Both the proton 7/2[523] and neutron 7/2[633] orbitals are intruder ones, therefore they are mainly concentrated on the spherical shell model vectors $|5 \ 5 \ 11/2 \ 7/2\rangle$ and $|6 \ 6 \ 13/2 \ 7/2\rangle$ respectively, although other vectors with smaller coefficients also contribute, as seen in Tab.~\ref{tab:3} and \ref{tab:4}.

\subsection{The nucleus $^{180}$Ta}
\label{sec:nucleusTa}

The even-even core of $^{180}_{~73}$Ta$_{107}$ is $^{178}_{~72}$Hf$_{106}$, for which the experimental value of the collective deformation variable $\beta$ is 0.2779~\cite{Pritychenko}, thus the Nilsson deformation $\epsilon=0.95 \beta$ \cite{NR} is 0.2640~.  

Several different theoretical calculations, including covariant density functional theory using the DDME2 functional~\cite{CDFTPLB,CDFTPRC},  Skyrmre-Hartree-Fock-BCS\footnote{see also N. Minkov, private communication.}~\cite{Minkov1}, as well as a two quasi-particle plus rotor model in the mean field represented by a deformed Woods-Saxon potential~\cite{Patial} agree that the first neutron orbital lying above the Fermi surface of the core nucleus $\rm ^{178}_{~72}Hf_{106}$ is the 9/2[624] orbital, while the first proton orbital lying above the Fermi surface of the core nucleus $\rm ^{178}_{~72}Hf_{106}$ is the 9/2[514] orbital. 
Therefore it is safe to assume that these two orbitals will play a major role in the formation of the  $9^-$ isomer state of $\rm ^{180}_{~73}Ta$. 

The question then comes from which orbitals the excited state $2^+$ may arise. 
The above mentioned covariant density functional theory using the DDME2 functional\footnote{K.E. Karakatsanis, private communication.}~\cite{CDFTPLB,CDFTPRC}, calculations and  Skyrmre-Hartree-Fock-BCS \cite{Minkov1} calculations indicate that the last neutron orbital below the Fermi surface is the 5/2[512] orbital, while the last proton orbital below the Fermi surface is the 7/2[404] orbital. 
Then it is plausible that the $2^+$ excited state will come from combining the proton 9/2[514] orbital (the first orbital above the Fermi surface) with the neutron 5/2[512] orbital (the first orbital below the Fermi surface).

It is instructive to consider the formation of the aforementioned states under the light of the expansions of the Nilsson orbitals in terms of spherical shell model orbitals, shown in Tab.~\ref{tab:3} and \ref{tab:4} 
The orbitals participating in the formation of the $9^-$ isomer, proton 9/2[514] and neutron 9/2[624], are both intruder orbitals, thus the main contribution comes from the $ |5 \ 5 \ 11/2 \ 9/2 \rangle$  component of the former and the $| 6 \ 6 \ 13/2 \ 9/2\rangle$ component of the latter. 
The orbitals participating in the formation of the $2^+$ excited state are the proton 9/2[514] (intruder) and neutron 5/2[512] (normal parity) orbitals, from which the leading contribution will come from the $ |5 \ 5 \ 11/2 \ 9/2\rangle$ and $| 5 \ 3 \ 7/2 \ 5/2\rangle$ vectors respectively.

\section{Considerations  of the  $\rm ^{166}Ho$ target}

So we will begin with the nucleus $^{166}$Ho, which is well studied, see, e.g.,~\cite{ISRR87}. This is an odd-odd nucleus (Z=67, N=99), which is a deformed nucleus described in the Nilsson model by  $ \left (\frac{7}{2}\right )^-[523]$ for protons, deformed level associated with the spherical $0h$ level,  and  the $ \left (\frac{7}{2}\right )^+[633]$ associated with the spherical level $0i$ for neutrons. It is convenient to express these states in a spherical basis in terms of a deformation~\cite{Casten90}. 
Thus for the purpose of our calculation it is sufficient to consider the expression
\begin{equation}
\left (\frac{7}{2}\right )^-[523]\leftrightarrow C_{0h_{11/2}}|0h_{11/2}\rangle,\,\left (\frac{7}{2}\right )^+[633]\leftrightarrow C_{0i_{11/2}}|0i_{11/2}\rangle+C_{0i_{13/2}}|0i_{13/2}\rangle
\end{equation}
with 
\begin{equation}
C_{0h_{11/2}}=0.9836,\,C_{0i_{11/2}}=-0.1,C_{0i_{13/2}}=0.9658
\end{equation}
It is reasonable to assume that this odd-odd nucleus can be considered as a two particle system composed of one proton and one neutron in the above levels. 
Thus we get 
\begin{equation}
\begin{split}
0^{-}&= C_{0i_{11/2}}C_{0h_{11/2}}\left [|0i_{11/2}\otimes 0h_{11/2}\rangle \right] ^{0}, \\
7^{-}&=C_{0h_{11/2}}\left ( C_{0i_{11/2}}\left [|0i_{11/2}\otimes 0h_{11/2}\rangle \right] ^{7}+|C_{0i_{13/2}}\left [ |0i_{13/2}\otimes 0h_{11/2}\rangle \right ] ^{7}\right )
\end{split}
\end{equation}

\subsection{The nuclear matrix elements}

In spite of the fact that  the coefficient $C_{0i_{11/2}}$ is small, the inclusion of the  $0i_{11/2}$ is mandatory to make the $0^-$ ground state wave function. The inclusion of the  $0i_{13/2}$
state with the large coupling is helpful but it can  not lead to proton  induced transitions. So one expects a suppression of NME. The operator for the transition has the structure 
$T^{\lambda,J}$ with rank $J=7$ and orbital rank $\lambda=J-1,J,J+1$. The interaction cannot convert protons to neutrons or vice versa. So the $\lambda=J=7$ is excluded by parity conservation.  Thus $\lambda=J=6,8$, i.e. only the spin induced transitions are allowed.
As a result  in this case the ratio of the NME divided  by the corresponding one for the nucleon can be cast  in the form:
\begin{equation}
R_{ME(q^2)}^2=\frac{|ME(q^2)|^2_{nuc}}{|ME_N|^2}=C_{VA} \left ( R_{ME(q^2)}^2\right )_0,\, C_{VA}=\frac{f_A^2}{f_V^2+3f_A^2}
\label{Eq:MEratio}
\end{equation}
The function $R_{ME(q^2)}^2$ is independent of the  scale, but it does depend on the ratio $f_A/f_V$ via the coefficient
\begin{equation}
C_{VA}=\left \{ \begin{array}{cc}~~0,& f_V >>fA\\1/3,& f_V <<fA\\1/4,& f_V \approx fA\\
\end{array}
\right .
\end{equation}
The vector current does not contribute in this case, but the first line in this expression comes from the explicit dependence of the cross section on the couplings (see Eq.\eqref{Eq:finalexpr}). 
In evaluating the NMEs one needs the reduced matrix element 
$$RME=\langle 0^6|| T^{\lambda,J}||7^{-}\rangle$$
Using standard  Racah techniques \cite{Hecht2000} one can obtain Eq.\eqref{Eq:FullRNME} of the Appendix~\ref{app:reduced_ME}. A detailed explicit calculation reduced matrix elements in the case of $\rm^{166}Ho$ is given in section~\ref{sec:pnHo}.

After that one can incorporate into  the reduced matrix element the form factor associated with in each orbit, obtained  via the corresponding radial integrals of the spherical Bessel  $j_{\lambda}(qr)$ finding this way the  single particle form factor for each orbit, see Appendix~\ref{app:shell_model}.

\begin{figure}[ht]
\centering
\includegraphics[width=0.7\textwidth]{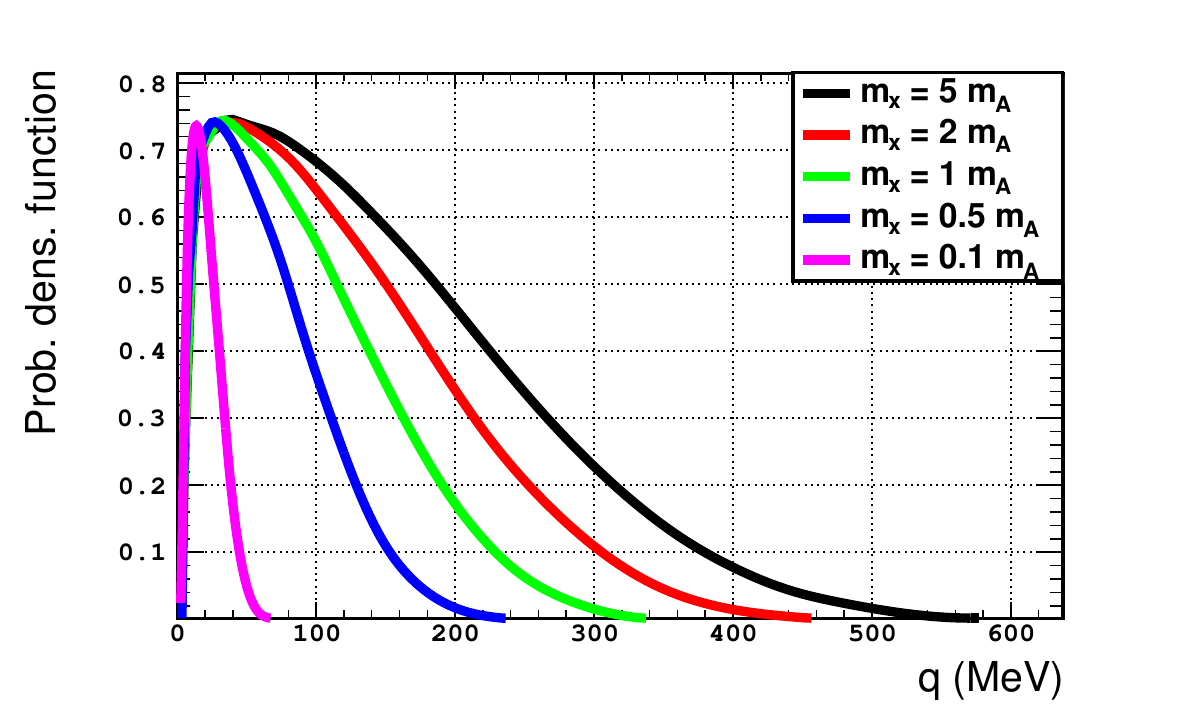}\\
\caption{The allowed momentum distribution (defined in Appendix~\ref{app:diff_xsec}) arising from the maximum allowed velocity (escape velocity) of the distribution for $\rm^{166}Ho$. Different line colors correspond to the WIMP masses $m_{\chi}=( 0.1,\,0.5,\,1,\, 2,\, 5)m_A$ }
\label{fig:psi(q)Ho}
\end{figure}

\begin{figure}[ht]
\centering
\includegraphics[width=0.7\textwidth]{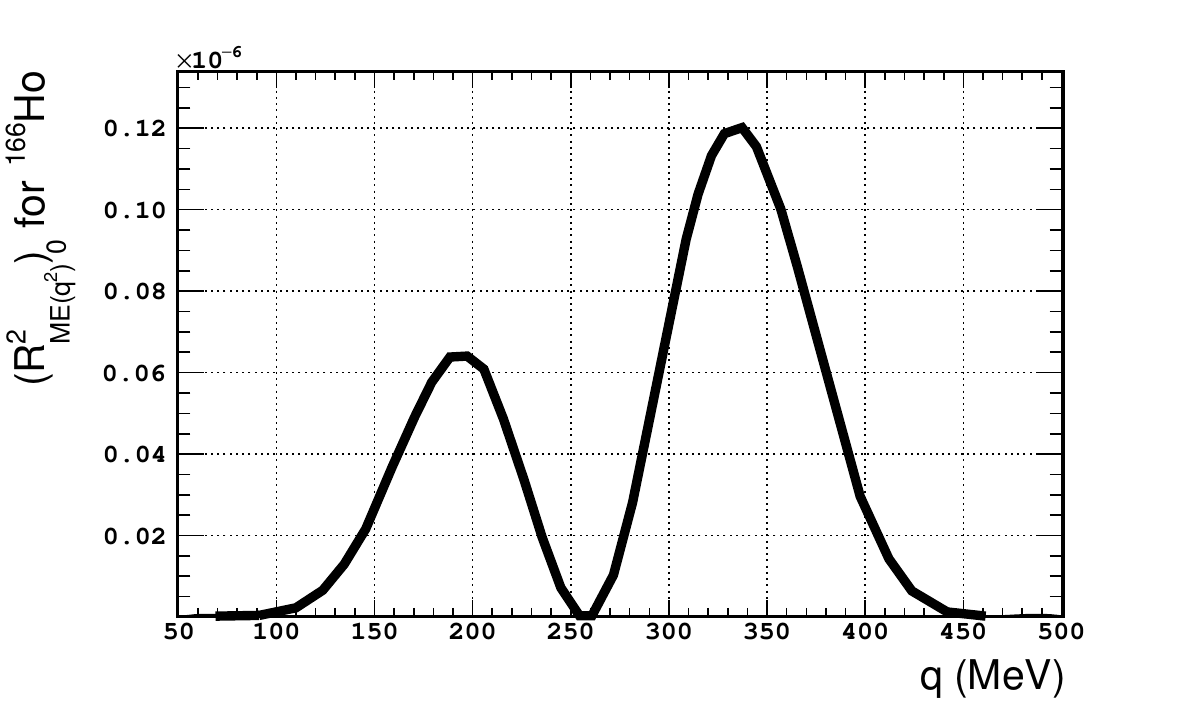}
\caption{The function $\left (R_{ME(q^2)}^2\right )_0$ for $^{166}$Ho, which  essentially is the ratio of the NME  divided by the corresponding one for the nucleon,  with the factor  $\frac{f_A^2}{3f_A^2+f_{V}^2}$ removed.  The part of the space above $q_{\rm max}=(65.6, 233, 349, 464, 579$ is not allowed for the case of $m_{\chi}=( 0.1,\,0.5,\,1,\, 2,\, 5)m_A$ respectively (see Fig. \ref{fig:psi(q)Ho}  and the text for details).}
\label{fig:FFRsqHo}
\end{figure}

\begin{figure}[ht]
\centering  
\includegraphics[width=0.7\textwidth]{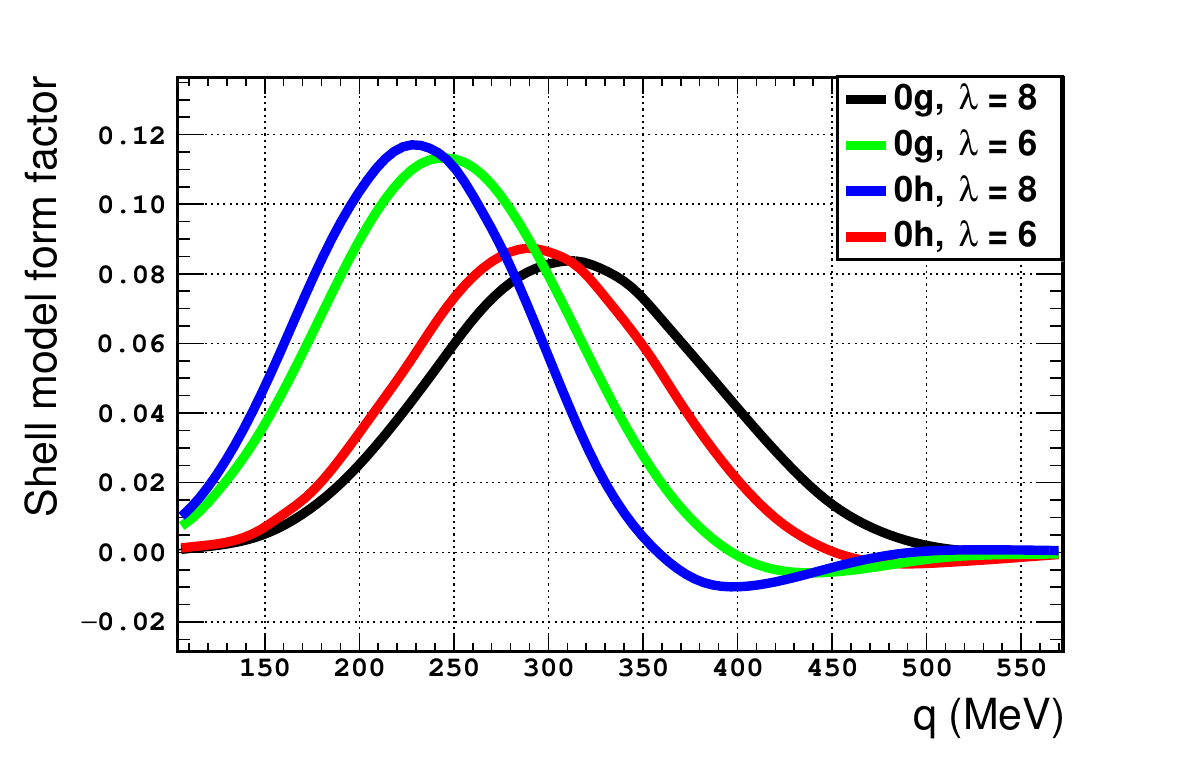}
\caption{The shell model single particle form factors  encountered in the case of a  WIMP induced nuclear transition with $\Delta J=7 $ for $^{166}$Ho, with $\lambda$ is the orbital rank of the operator (multi-polarity).}
\label{fig:FFsmHo}
\end{figure}

\begin{figure}
     \centering
     \begin{subfigure}[b]{0.49\textwidth}
         \centering
         \includegraphics[width=1.1\textwidth]{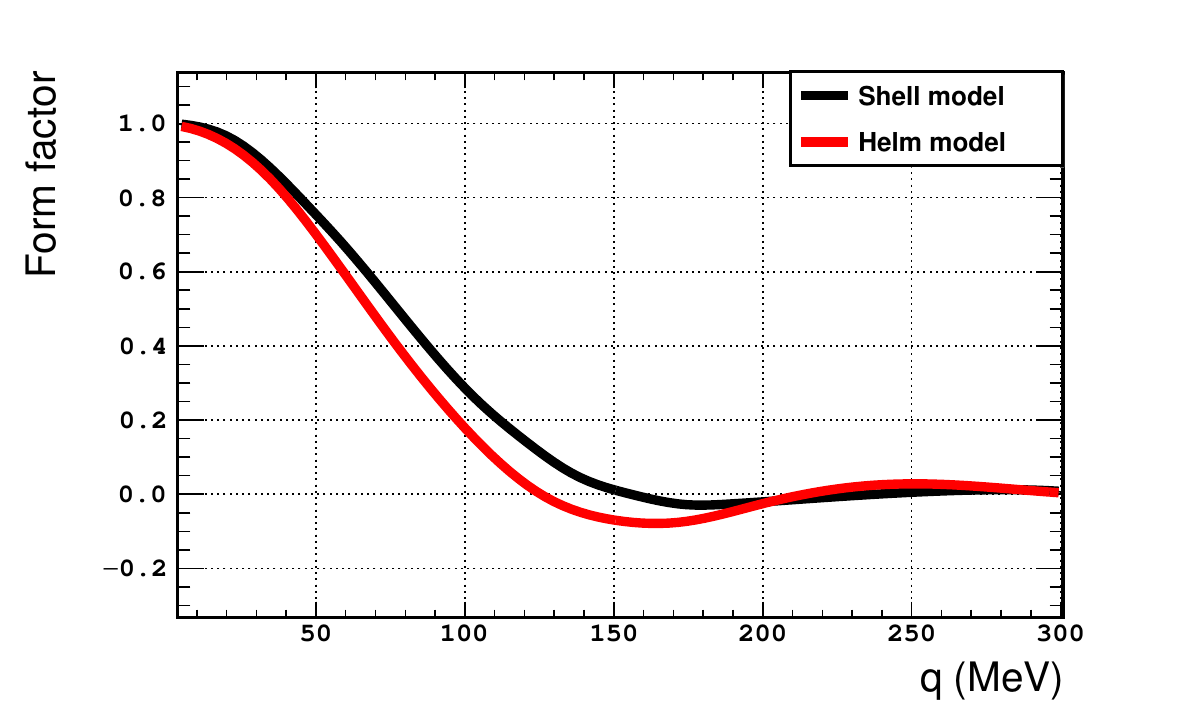}
         \caption{}
         \label{fig:FFelHo}
     \end{subfigure}
     \hfill
     \begin{subfigure}[b]{0.49\textwidth}
         \centering
         \includegraphics[width=1.1\textwidth]{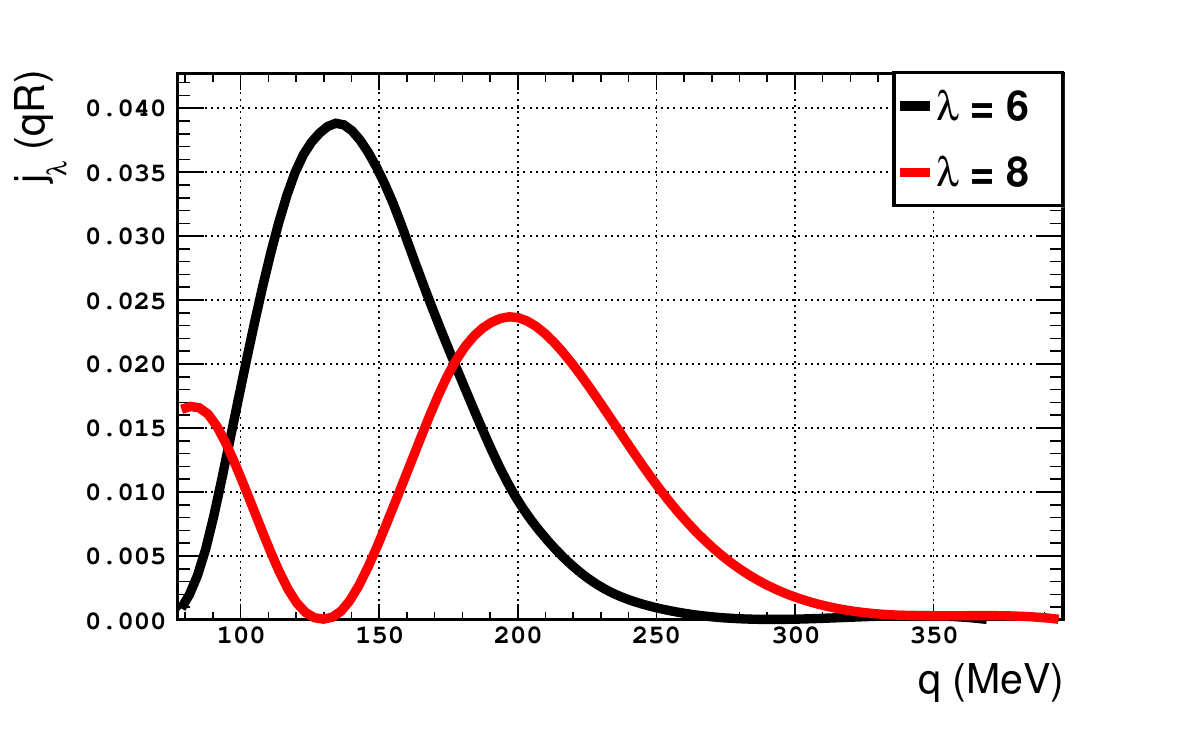}
         \caption{}
         \label{fig:j6and8}
     \end{subfigure}
\caption{(a)~The shell model  form factor  encountered in the case of a  WIMP induced elastic nuclear transition for $^{166}$Ho and the same with the Helm form factor.
The NME in this case is obtained by multiplying this form factor with the number of nucleons in the nucleus, here $A=166$.
(b)~The function $j_{\lambda}(q R)$ for an A=166 nucleus in the range of $q$ of interest in the present work for the appropriate values of $\lambda$.}
\end{figure}

Let us now consider   the allowed range of the momentum transfer. As we can see the minimum velocity must be smaller than the escape velocity, see Eq.\ref{Eq.vminofq}. For a  M-B distribution 
see Eq.(\ref{Eq:yoferf}) with $y_i=\upsilon_i / \upsilon_0,~i=1,2$. 
This momentum dependence is exhibited in Fig.~\ref{fig:psi(q)Ho}.
The NME  $\left (R_{ME(q^2)}^2\right )_0$ is exhibited in Fig.~\ref{fig:FFRsqHo}. We observe that it is greatly suppressed.
This may be  surprising in view of the fact that all single particle form factors involved are  much larger, see Fig.~\ref{fig:FFsmHo} and take its square. 
These form factors, however, are much smaller the  nuclear form factors encountered in the case of the standard  WIMP searches involving the elastic WIMP-nucleus scattering, see Fig .~\ref{fig:FFelHo}.

This suppression appears to be mainly due to the geometric factors involved in the reduced matrix elements, i.e.  nine-j symbols, Racah functions etc as well as  due to smallness of the coefficient $C{i_{11/2}}$. Indeed  the total NME can be written as
$$ NME_t= \sum_{\ell,\lambda}C_{\ell,\lambda}FF_{\ell,\lambda} (q)$$
From the above reduced ME we find 
\begin{equation}
C_{0i,8}=0.0203,\,C{_0i,6}=-0.01935,\,C_{0f,8}=0.000737,\,C_{0f,6}=0.00374 .
\label{Eq:CSM}
\end{equation}
  The corresponding shell model single particle form factors are exhibited in Fig. ~\ref{fig:FFsmHo}. There appears to be some cancellation due to their size, but it is mainly a consequence of the fact that all the coefficients $ C_{\ell,\lambda}$ are small and not of the same sign. Furthermore recall that, after incorporating the statistical factor   
  $$\left (R_{ME(q^2)}^2\right )_0=\frac{1}{15}( NME_t)^2$$
  
In addition  it is also likely that the single particle form factors in the shell model are suppressed, or at least much smaller than implied by the scale set by the spherical Bessel functions $j_{\lambda}$, see Fig.~\ref{fig:j6and8}. 
This suppression is not very significant, but  ideally all fours single particle form factors could be determined experimentally.

\begin{figure}[ht]
\centering 
\includegraphics[width=0.7\textwidth]{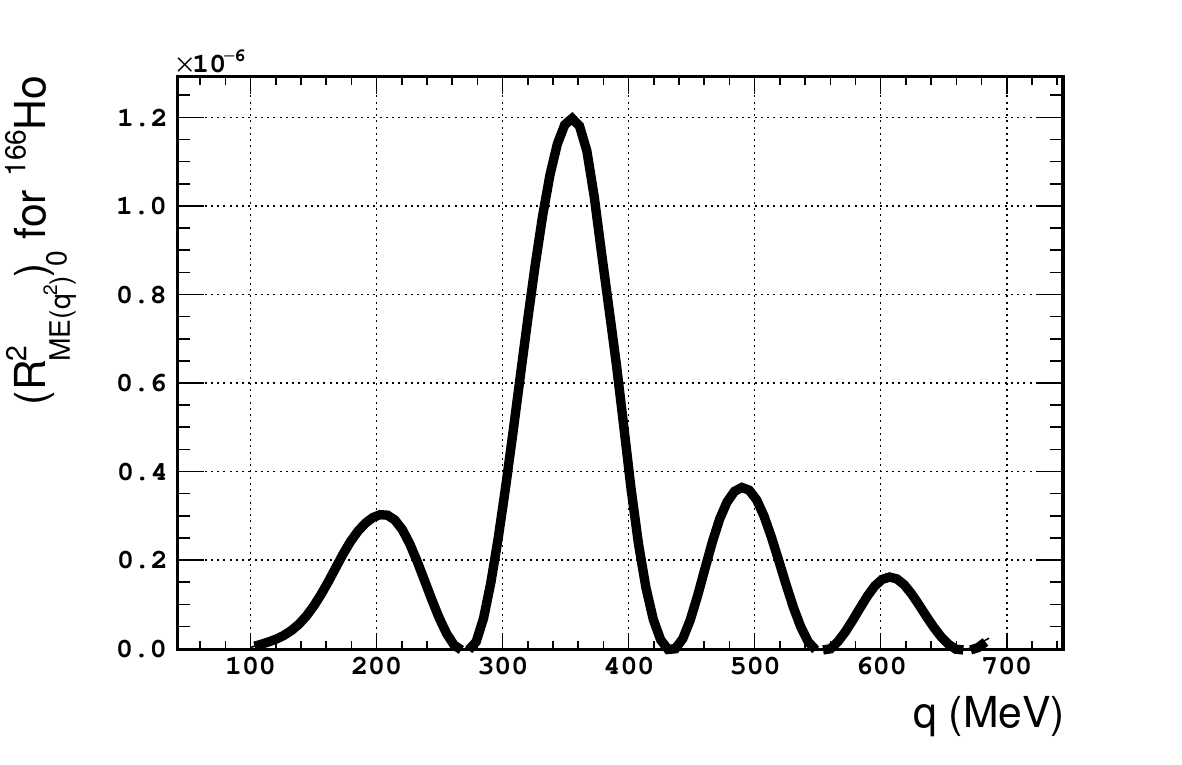}
\caption{The same as in Fig. 	\ref{fig:FFRsqHo} obtained with Helm type single particle form factors.	The restrictions on the allowed momenta are the same as in Fig.~\ref{fig:FFRsqHo}.
}
\label{fig:FHRsqHo}
\end{figure}

\begin{figure}
\centering 
\includegraphics[width=0.7\textwidth]{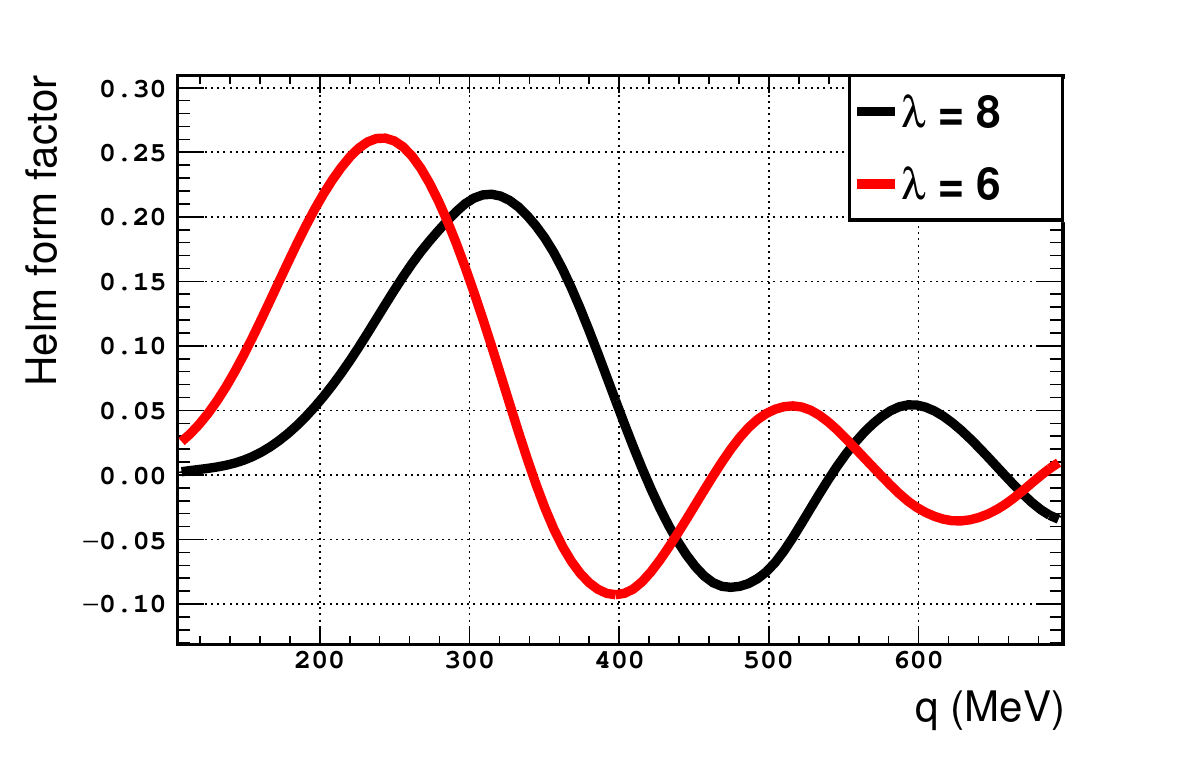}
\caption{The Helm form factors $FFH_{\lambda}(q)$ for $\lambda=6$,  red line, and for $\lambda=8$,  black line, in the case of $^{166}$Ho and momentum transfer of interest in the present work.}
\label{fig:FFRHo}
\end{figure}

For orientation purposes we will consider Helm like single particle form factors:
\begin{equation}
 F_{\lambda}(q)=(2 \lambda +1) e^{-\frac{1}{2} a^2 q^2}	\frac{j_{\lambda }(q R)}{q R}
 \label{Eq:HelFF}
\end{equation}
 This is an extension of the Helm-form factor \cite{Helm56}, normalized so that it is unity for $\lambda=1$ at $q=0$. 
This expression for $a=0.6$F and $R=6.87$F is exhibited in Fig.~\ref{fig:FFRHo}.
The NME using these form factors is exhibited in Fig.~\ref{fig:FHRsqHo}.

There is now an improvement of two orders of magnitude, but the obtained result is still  quite small. 

In the case of the Helm-like form factors proceeding as in the case of shell model one finds
\begin{equation}
RME(q)=\sum_{\lambda}C_{\lambda}FFH_{\lambda}(q)
\end{equation}
with
\begin{equation}
 C_{\lambda}={0.02107, -0.01561}
 \label{Eq:CHelm}
\end{equation}
 These coefficients are also small.  
 The negative sign in Eqs.\eqref{Eq:CSM} and \eqref{Eq:CHelm}  leads to suppression of the NME.

\subsection{Some results for $\rm ^{166}$Ho}

\begin{figure}
     \centering
     \begin{subfigure}[b]{0.49\textwidth}
         \centering
         \includegraphics[width=1.1\textwidth]{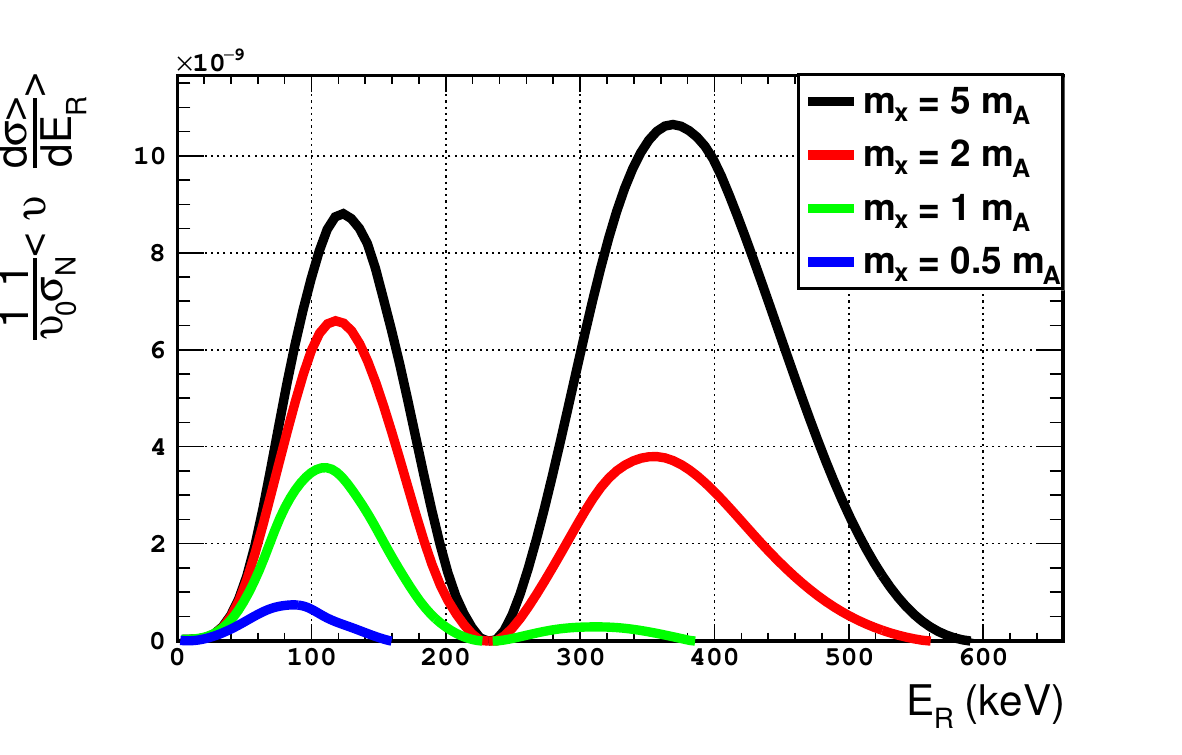}
         \caption{}
         \label{fig:funsv}
     \end{subfigure}
     \hfill
     \begin{subfigure}[b]{0.49\textwidth}
         \centering
         \includegraphics[width=1.1\textwidth]{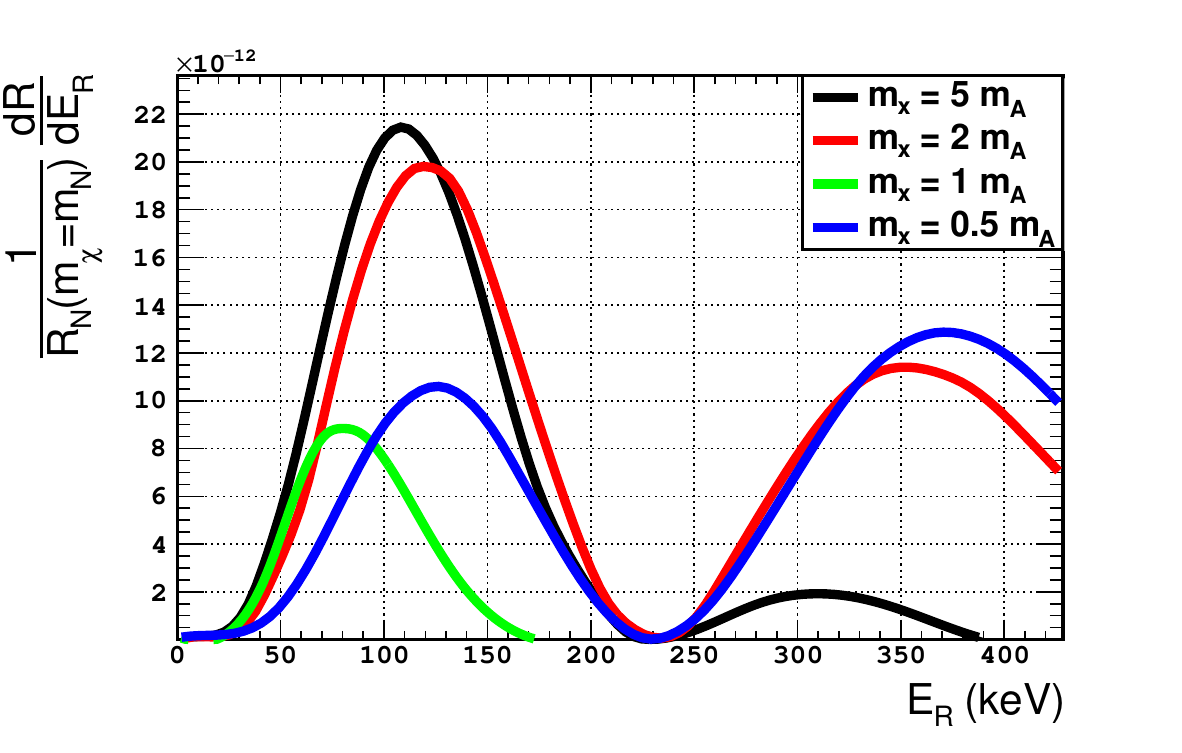}
         \caption{}
         \label{fig:DifRatesv}
     \end{subfigure}
\caption{(a)~The function $\frac{1}{\upsilon_0}\frac{1}{\sigma_N}\langle \upsilon  \frac{d\sigma>}{dE_R}\rangle$ in units of keV$^{-1}$.  
The Helm type form factor has been employed.
(b)~The differential rate relative to the total nucleon rate (for $m_{\chi}=m_N$),  $\frac{1}{R_N(m_{\chi}=m_N)} \frac{dR}{dE_R}$, in units of keV$^{-1}$.  
The Helm type form factor has been employed.
It also  contains the WIMP mass dependence arising from the WIMP density in our galaxy.}
\end{figure}

The numerical value of  $$\Lambda \frac{m_A}{m_N^2}\frac{1}{\upsilon^2_0}\frac{1}{2 \pi} $$ 
in Eq.\ref{Eq:finalexpr} using Eq. \eqref{Eq:MEratio} with $f_A/f_V=1$, is 0.063 for A=166, expressed in units of keV$^{-1}$. 
The plot for  $\frac{1}{\upsilon_0}\frac{1}{\sigma_N}\Big\langle \upsilon  \frac{d\sigma}{dE_R}\Big\rangle$ 
versus the previous one multiplied with $0.063$. 
We prefer to express it as a function of $E_R$ in units of keV, i.e Fig.~\ref{fig:funsv}. 

The expressions for $\sigma_N$, $\phi$ and $R$ can be obtained using the relevant values for the nucleon:
$$\sigma_N=3.5 \cdot 10^{-39}(f_V^2+3 f_A^2), {\rm\qquad cm^2}$$
$$\Phi=2.1 \cdot 10^{38} \frac{m_N}{m_\chi}, \qquad  {\rm cm^{-2}y^{-1}}$$
(kinematics factor), yielding (see the Appendix~\ref{app:NWimpRate}):\\
$$R_N=\Phi\sigma_N N_N=0.72\cdot(f_V^2+3 f_A^2), \qquad{\rm y^{-1}}$$


The same result holds for the obtained differential rate $$\frac{1}{R_N(m_{\chi}) }\frac{dR}{dE_R}$$ since the WIMP density used in obtaining the  densities is the same. 
The situation is, however, different if one is comparing the obtained differential rate relative to the total rate for the nucleon at some fixed value of the WIMP mass.
Indeed the obtained differential rate relative to  total rate of the nucleon for $m_N/m_{\chi}=1$ is exhibited in Fig.~\ref{fig:DifRatesv}. The exhibited differential rate  contains, of course,  the WIMP mass dependence arising from the WIMP density in our galaxy.


With such results the WIMP detection with the $\rm^{166}Ho$ appears very problematic.

\section{Considerations  of the $\rm ^{180}Ta$ target}

\begin{figure}
\centering
\includegraphics[width=0.7\textwidth]{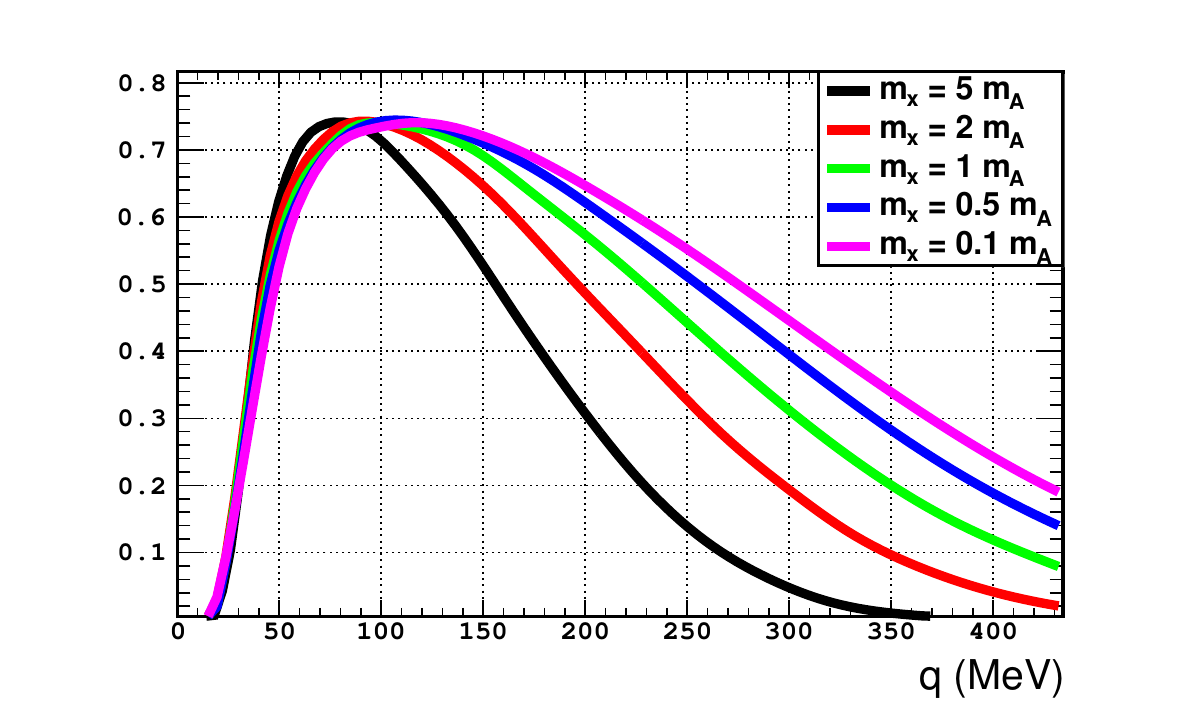}\\
\caption{The allowed momentum distribution (defined in Appendix~\ref{app:diff_xsec}) arising from the maximum allowed velocity (escape velocity) of the distribution, in the case of $\rm^{180}Ta$.
For different WIMP masses $m_{\chi}=( 0.1,\,0.5,\,1,\, 2,\, 5)m_A$. The transition energy is $\Delta=37$ keV.}
\label{fig:psi(q)Ta}
\end{figure}

We begin by considering the transition of the isomeric $9^-$ state to the $2^+$ state. The momentum dependence of the cross section arising from the velocity distribution is different from that of $\rm^{166}Ho$, since the transition energy is $\Delta=37$ keV. Thus the analog of Fig. ~\ref{fig:psi(q)Ho} is given in Fig.~\ref{fig:psi(q)Ta}.

To proceed further we need to determine the structure of  the target $\rm ^{180}Ta$. 
As explained in section \ref{sec:nucleusTa} in the context of the Nilsson model we can consider the proton  orbital $\frac{9}{2}$[514] both in the initial state $9^{-}$ and the final $2^+$. Furthermore for the neutrons we use  $\frac{9}{2}$[624] for the $9^-$ and the $\frac{5}{2}$[512] for the $2^+$. To proceed further we use the expansion of the Nilsson orbitals into shell model states found  in Tables~\ref{tab:3} and \ref{tab:4} for deformation parameter 0.30. Note that in this case only the neutrons can undergo transitions, while the protons are just spectators. In the case of the shell model one encounters 8 transition types with odd multi-polarities.

\subsection{Shell model form factors}

\begin{figure}
\centering
    \begin{subfigure}[b]{0.49\textwidth}
         \centering
         \includegraphics[width=1.1\textwidth]{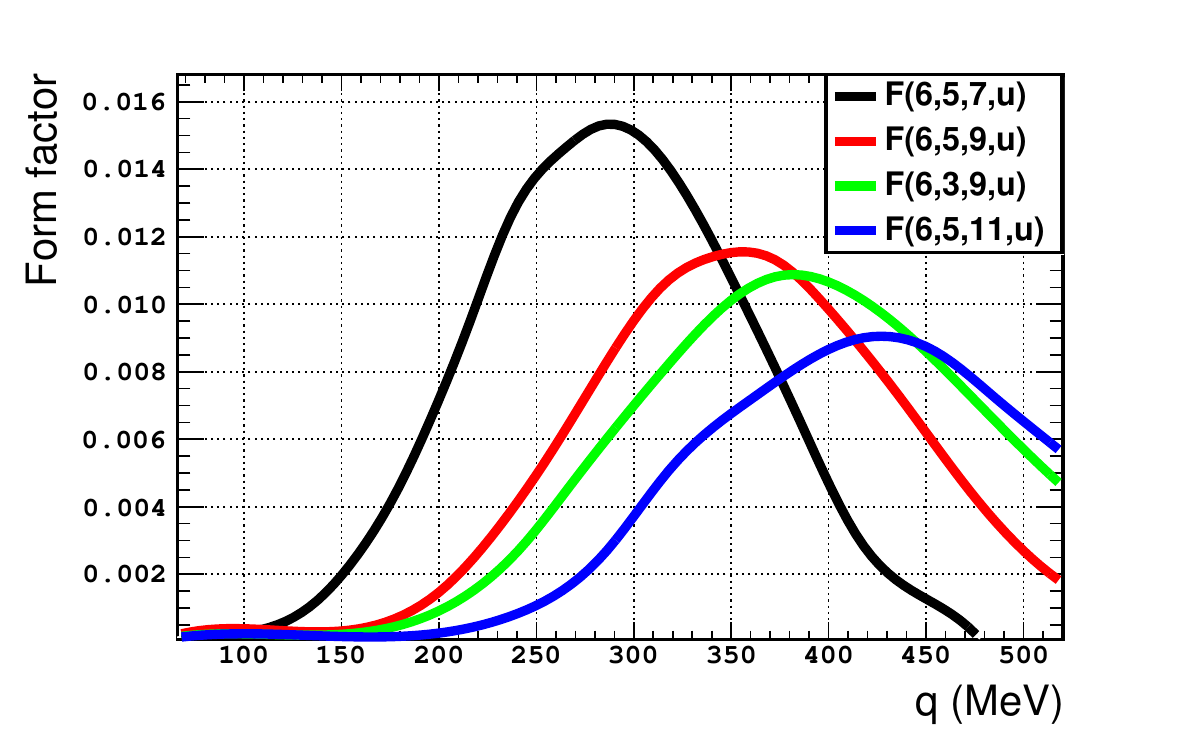}
         \caption{for F(6,5,7,u), F(6,5,9,u), F(6,5,11,u) and F(6,3,9,u)}
         \label{fig:SMFFTa_a}
     \end{subfigure}
\hfill
    \begin{subfigure}[b]{0.49\textwidth}
         \centering
         \includegraphics[width=1.1\textwidth]{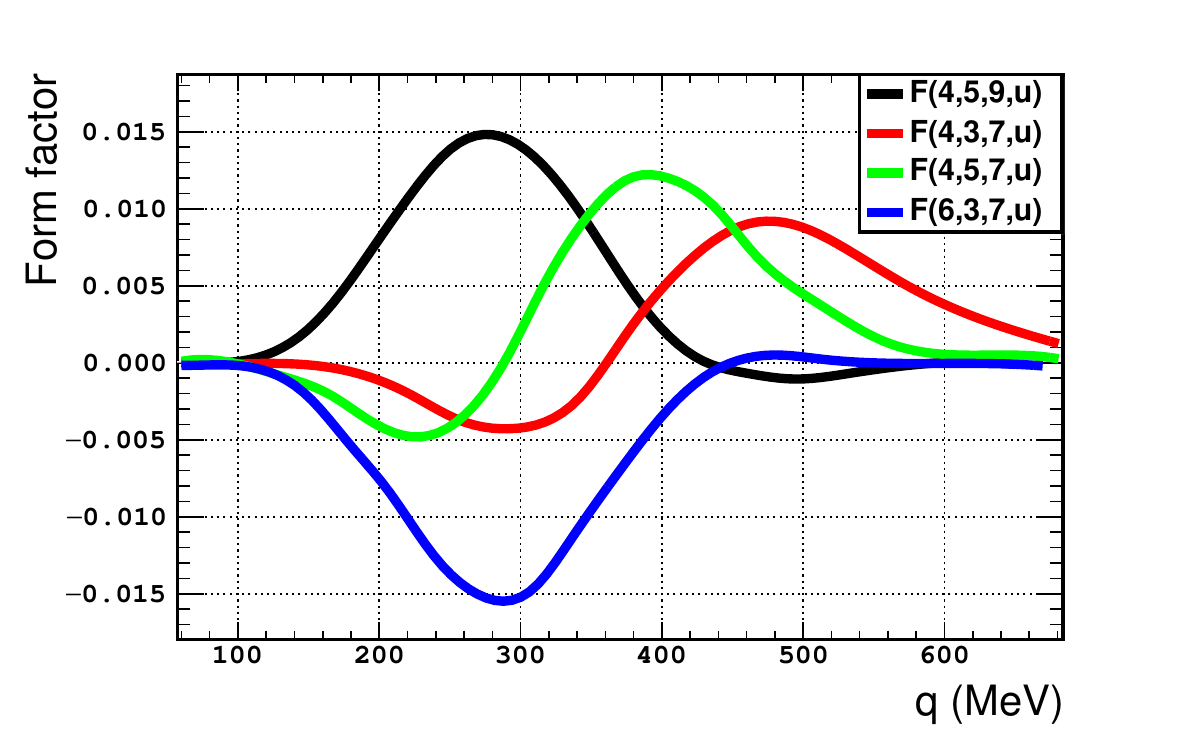}
         \caption{for F(6,3,7,u), F(4,5,7,u), F(4,5,9,u) and  F(4,3,7,u)}
         \label{fig:SMFFTa_b}
     \end{subfigure}
\caption{The form factors for different $F$ are exhibited.}
\label{fig:SMFFTa}
\end{figure}

The vector and axial vector reduced NMEs can be obtained using Eq.\eqref{Eq:PartialRNME} where the quantities with subscript 1 indicate neutrons and those with 2 are associated with protons. Thus we find:
$$RME_V=\frac{f_V}{f_A}(0.0644445
	F(4,3,7,u)+1.01419
	F(4,5,7,u)+1.01419
	F(4,5,9,u)+1.52946
	F(6,3,7,u)+$$ $$1.52946
	F(6,3,9,u)+1.52946
	F(6,5,7,u)+1.79799
	F(6,5,9,u)+2.19718
	F(6,5,11,u))$$
$$RME_A=0.321503 F(4,3,7,u)+2.05117
F(4,5,7,u)+2.16715
F(4,5,9,u)+2.04512
F(6,3,7,u)+3.3217$$ $$
F(6,3,9,u)+2.04512
F(6,5,7,u)+2.31181
F(6,5,9,u)+3.58938
F(6,5,11,u) $$ 
In the above expressions $F(\ell,\ell',\lambda)$ are the single particle form factors. The first two integers  indicate orbital angular momentum quantum numbers $\ell,\ell'$, while the last integer  $\lambda$ gives the multipolarity  of the transition. The quantity $u$ corresponds to $b_N q$, where $b_N$ is the harmonic oscillator length parameter. 
The NMEs have been normalized as above in the case $\rm^{166}Ho$, see Eq.\eqref{Eq:MEratio}, with a compensating factor of $f_A$ appearing explicitly  in the cross-section, Eq.\ref{Eq:finalexpr}. 
The relevant form factors are exhibited in Fig.~\ref{fig:SMFFTa}.


\begin{figure}
     \centering
     \begin{subfigure}[b]{0.43\textwidth}
         \centering
         \includegraphics[width=1.24\textwidth]{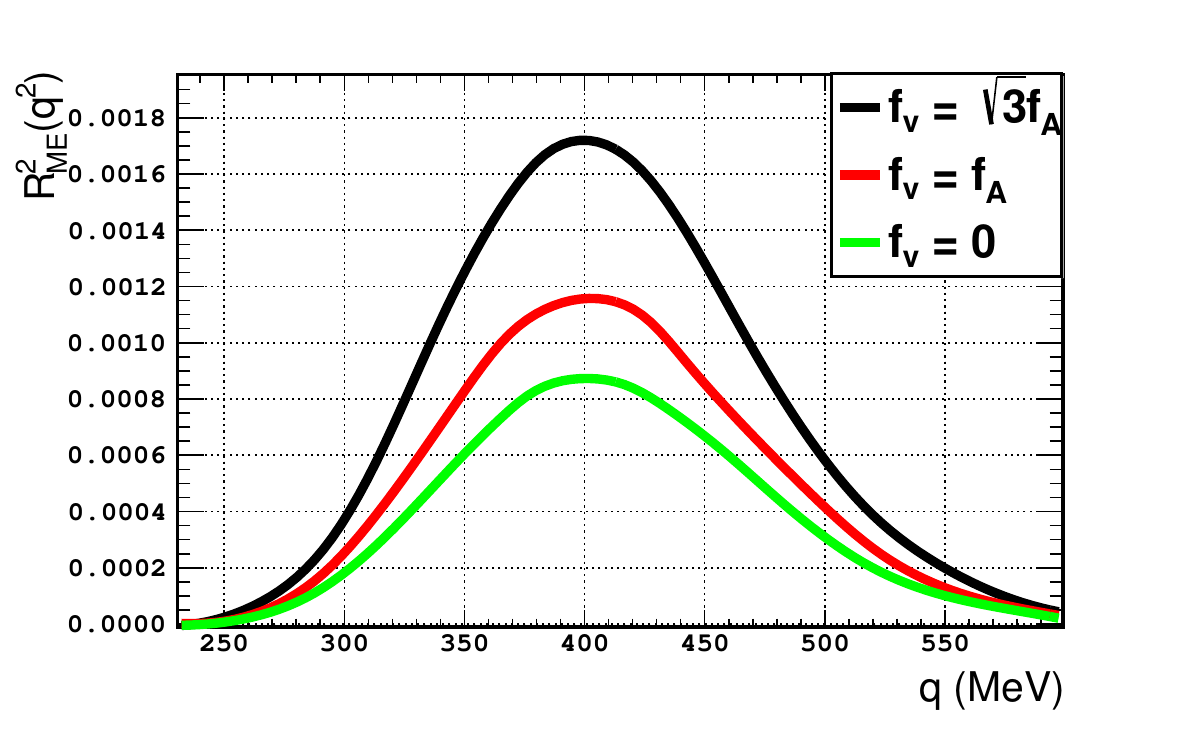}
         \caption{}
         \label{fig:SMMETa}
     \end{subfigure}
     \hspace{35 pt}
     \begin{subfigure}[b]{0.42\textwidth}
         \centering
         \includegraphics[width=1.15\textwidth]{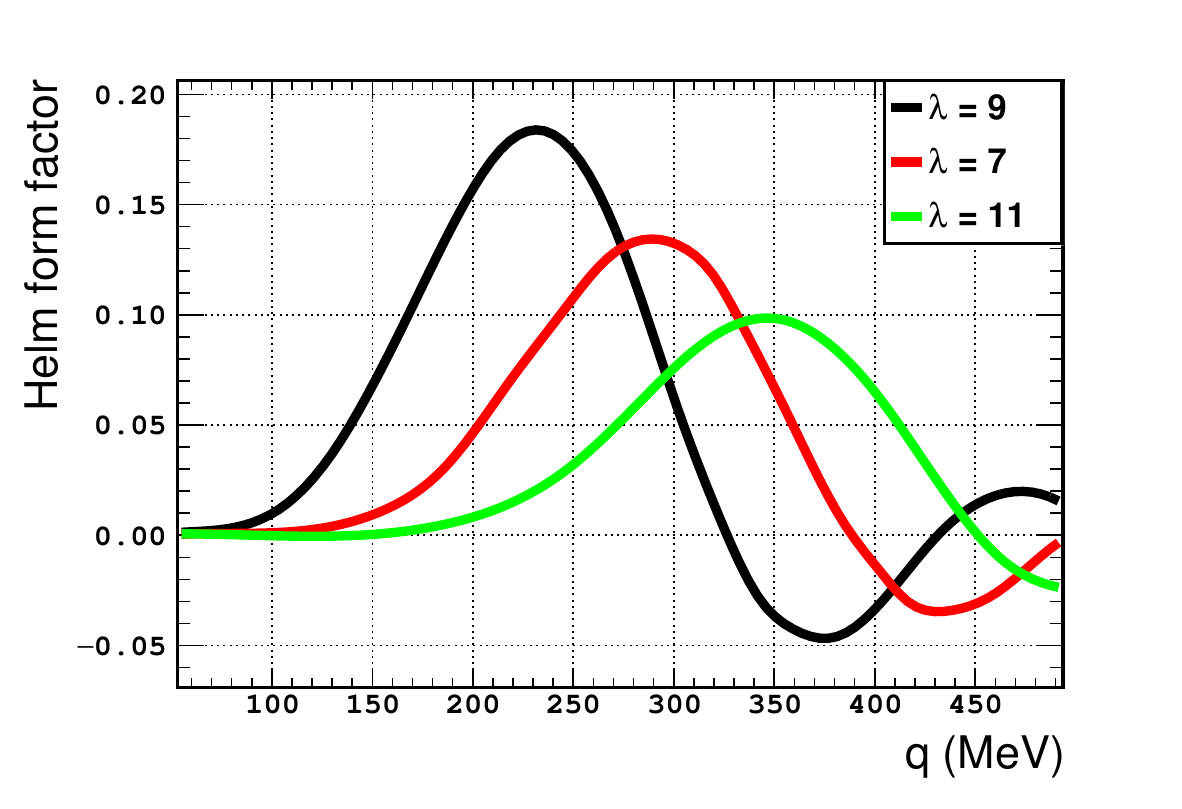}
         \caption{}
         \label{fig:HelmFFTa}
     \end{subfigure}
     \vfill
     \begin{subfigure}[b]{0.42\textwidth}
         \centering
         \includegraphics[width=1.15\textwidth]{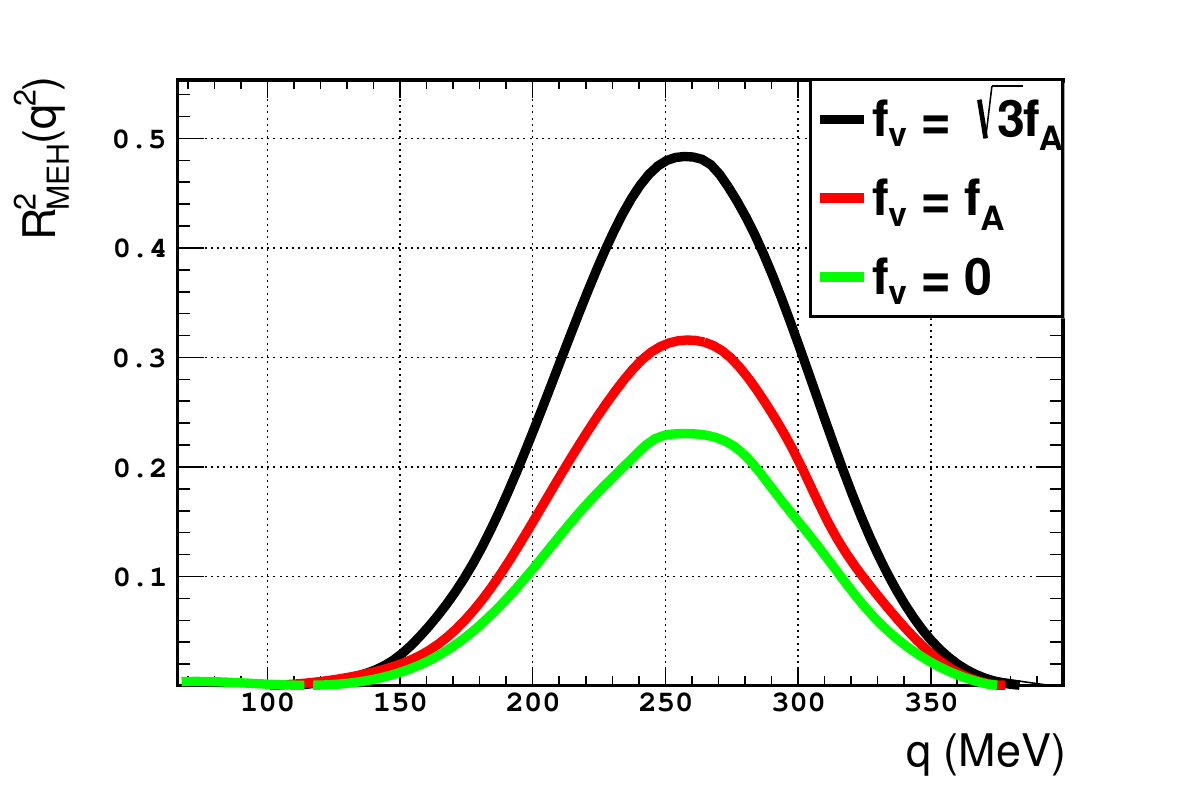}
         \caption{}
         \label{fig:HelmMETa}
     \end{subfigure}
\caption{(a)~The momentum dependence of the expression  $R^2_{ME}(q^2)$ is exhibited as a function of $q$ for different values of $f_V$.
(b)~The Helm type form factors for $\lambda=7$,  $\lambda=9$ and  $\lambda=11$.
(c)~The momentum dependence of $R^2_{MEH}(q^2)$ as a function of $q$ for different values of $f_V$.
Helm form factors have been used.}
\end{figure}

The relevant nuclear ME is given by:
\begin{equation}
R^2_{ME}(q^2)=\frac{1}{19}\left ( RME_V^2+RME_A^2\right )
\end{equation}
Its momentum dependence is exhibited in Fig. ~\ref{fig:SMMETa}. 
We should note that the large value  of the matrix element in the case of large $f_V$ is due to the normalization adopted to make the matrix element independent of the scale. 
Recall that the corresponding factor appears in the cross section. 
In the present work we will adopt $f_V =f_A$.


\subsection{Phenomenological  form factors}

It is generally believed that the shell model  single particle factors lead to large suppression,
so phenomenological form factors may be preferred. One example is the the Helm form factor, see Eq.\eqref{Eq:HelFF}. This has already been employed in the case of even transitions. 
We will employ here for odd (parity changing) transitions. 
Our treatment means that the radial integrals are independent of the angular momentum quantum numbers $\ell,~\ell'$. The obtained results are exhibited in Fig.~\ref{fig:HelmFFTa}.
The reduced  matrix elements for the vector  and the axial vector are:
\begin{equation}
\begin{split}
RMEH_A&=3.58938 F_{11}(a,q,R)+6.46292 F_7+7.80066 F_{9}(a,q,R) \\
RMEH_V&=\frac{f_V}{f_A}(2.19718 F_{11}(a,q,R)+4.13756 F_7(a,q,R)+4.34165 F_9(a,q,R)),
\end{split}
\label{Eq:coefNilsson}
\end{equation}
where $F_{\lambda}$ are the Helm single particle form factors. The NME is:
\begin{equation}
R^2_{MEH}(q^2)=\frac{1}{19}\left ( RMEH_V^2+RMEH_A^2\right )
\end{equation}
The momentum dependence of this ME is exhibited in Fig.~\ref{fig:HelmMETa}

\subsection{Some results for $^{180}$Ta}

\begin{figure}
     \centering
     \begin{subfigure}[b]{0.49\textwidth}
         \centering
         \includegraphics[width=1.06\textwidth]{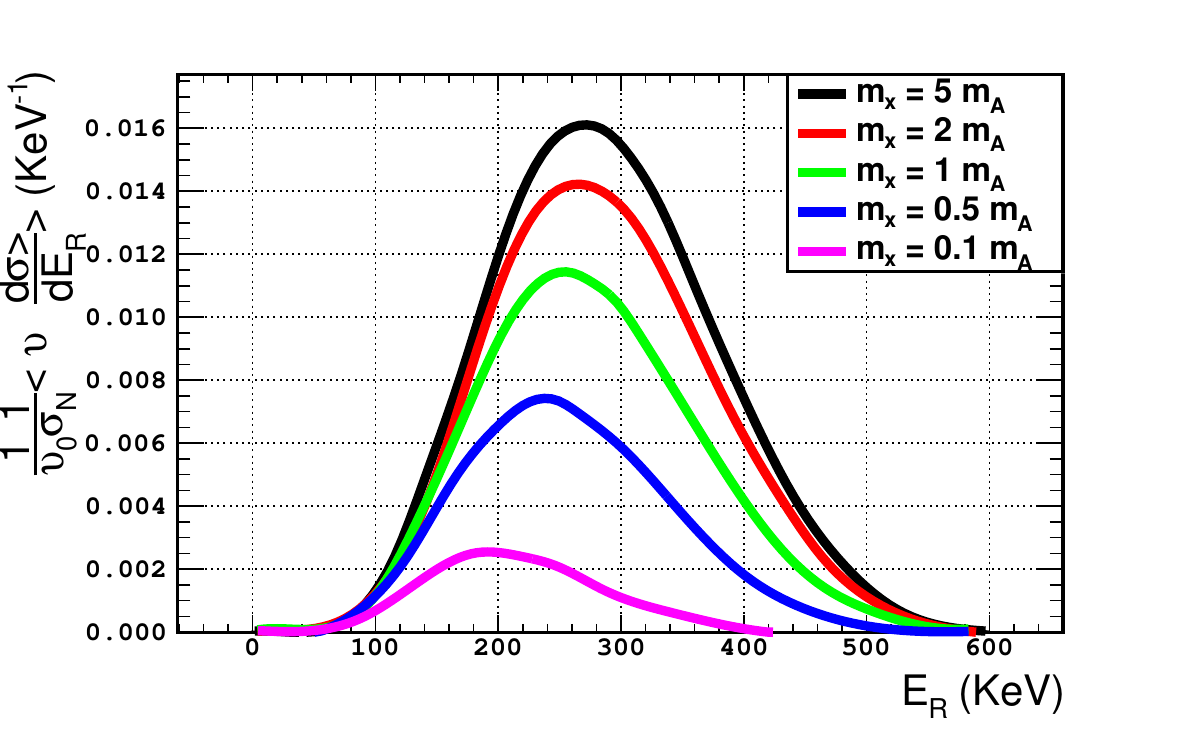}
         \caption{}
         \label{fig:funsvTa}
     \end{subfigure}
     \hfill
     \begin{subfigure}[b]{0.49\textwidth}
         \centering
         \includegraphics[width=1.18\textwidth]{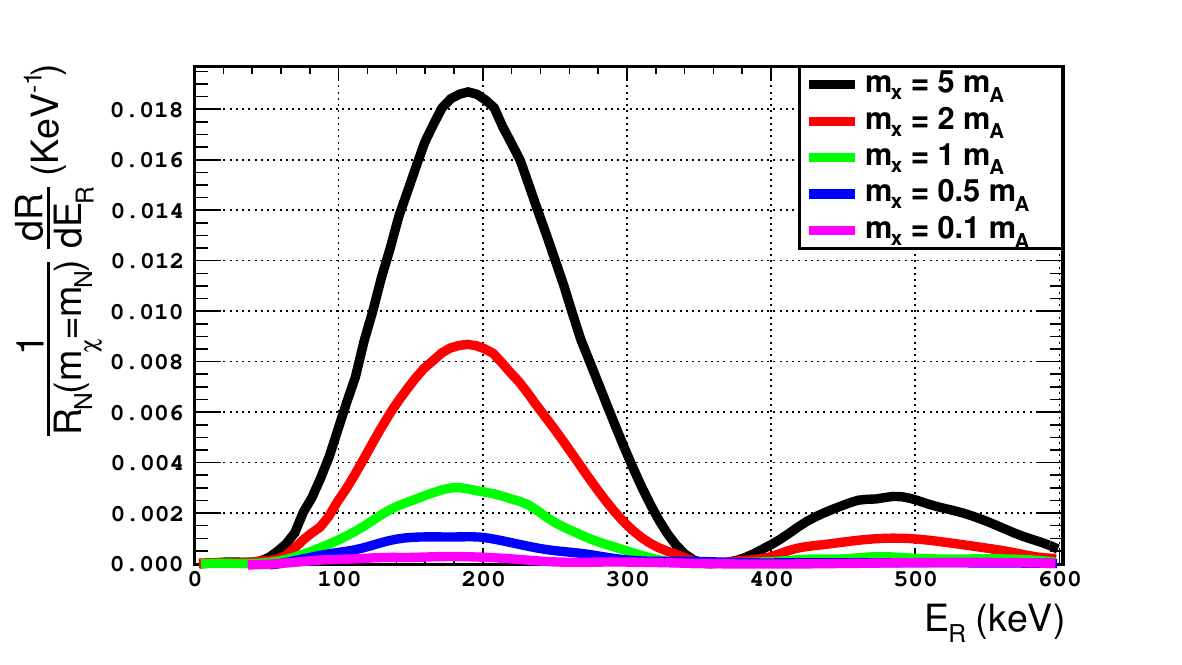}
         \caption{}
         \label{fig:DifRatesvTa}
     \end{subfigure}
\caption{(a)~The function $\frac{1}{\upsilon_0}\frac{1}{\sigma_N}\langle \upsilon  \frac{d\sigma>}{dE_R}\rangle$ in units of keV$^{-1}$. 
The Helm type form factor has been employed.
(b)~The differential rate relative to the total nucleon rate (for $m_{\chi}=m_N$),  $\frac{1}{R_N(m_{\chi}=m_N)} \frac{dR}{dE_R}$, in units of keV$^{-1}$ for the Ta target. The black curve in the drawing has been reduced by a factor of 5, so the related rate must be multiplied by 5.  The Helm type form factor has been employed. It also  contains the WIMP mass dependence arising from the WIMP density in our galaxy.}
\end{figure}

The numerical value of  $$\Lambda \frac{m_A}{m_N^2}\frac{1}{\upsilon^2_0}\frac{1}{2 \pi}$$ in Eq.\eqref{Eq:finalexpr} using Eq.\eqref{Eq:MEratio} with $f_A/f_V=1$, is 0.068 for A=180, expressed in units of keV$^{-1}$. 
The plot for 
$ \frac{1}{\upsilon_0}\frac{1}{\sigma_N}\big\langle \upsilon  \frac{d\sigma>}{dE_R}\big\rangle$
versus the previous one multiplied with 0.063. 
The preference is to express it as a function of $E_R$ in units of keV, i.e Fig.~\ref{fig:funsvTa}. 
It can be shown that a  similar expression holds for the rate 
$$\frac{1}{R_N}\frac{dR}{dE_R},$$ 
see Fig.~\ref{fig:DifRatesvTa}.
The expressions for $\sigma_N$ and $R$ can be obtained using the relevant values for the nucleon:
$$\sigma_N=8.8 \cdot 10^{-40}(f_V^2+3 f_A^2), {\rm\qquad cm^2}$$
$$R_N=\Phi\sigma_N N_N=0.72\cdot(f_V^2+3 f_A^2), \qquad{\rm y^{-1}}$$

\begin{figure}
\centering
\includegraphics[width=0.7\textwidth]{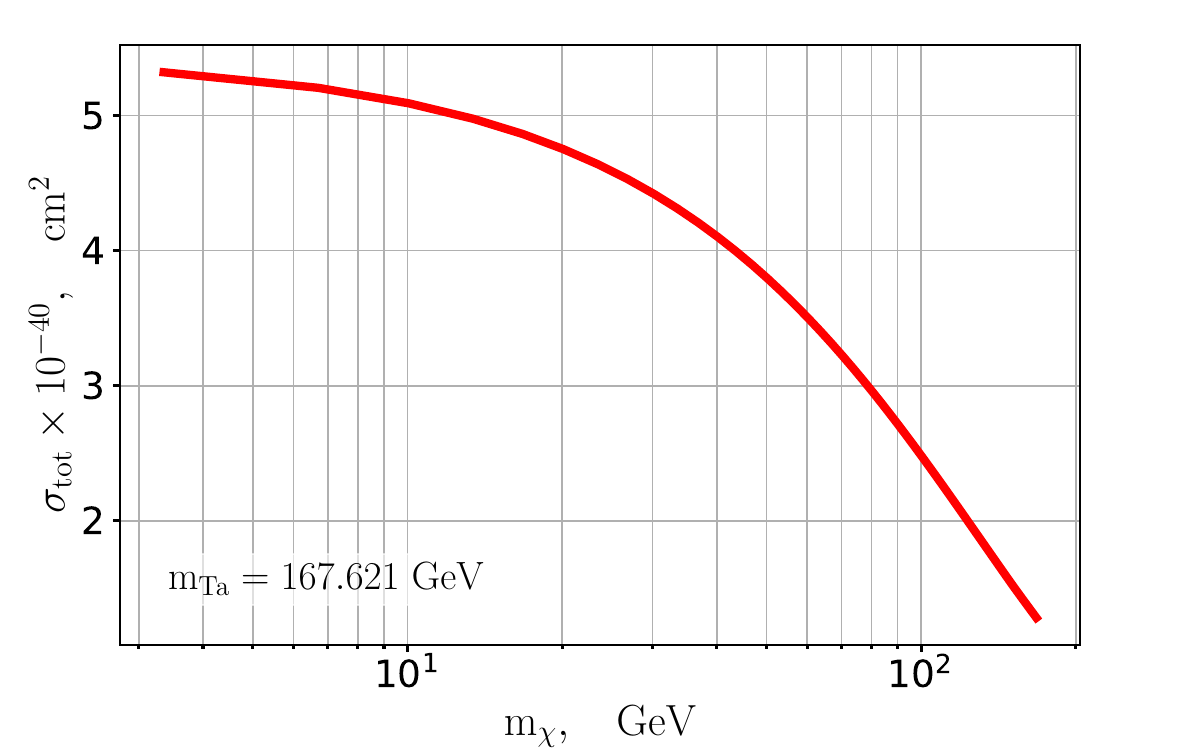}
\caption{The total WIMP-Nucleus cross section as a function of WIMP mass for the target nucleus $\rm ^{180m}Ta$.}   
\label{fig:Total_Xsec}
\end{figure}

One can integrate the differential cross section over the recoil energy $E_R$ and multiply with the total nucleus mass to obtain the WIMP-Nucleus cross section as a function of the WIMP mass $m_{\chi}$ this is exhibited in Fig. ~\ref{fig:Total_Xsec}.
\textcolor{black}{The dominant source of uncertainties for the cross section is the NME.
The multipolarities of high order and the momentum transfer introduce around a 30\% error. 
The escape velocity contributes 10\%.
In addition, there is an uncertainty in the model parameters given in Tabs.~\ref{tab:3},~\ref{tab:4} of about 10\%.
This, in turn, gives the total uncertainty as the square root of the sum of quadratures, approximately $33.2 \%$.}

The same result holds for the obtained differential rate $$\frac{1}{R_N(m_{\chi}) }\frac{dR}{dE_R}$$ since the WIMP density used in obtaining the  densities is the same. 
The situation is, however, changed if one is comparing the obtained differential rate relative to the total rate for the nucleon at some fixed value of the WIMP mass.
Indeed the obtained differential rate relative to  total rate of the nucleon for $m_N/m_{\chi}=1$ is exhibited in Fig.~\ref{fig:DifRatesvTa}. 
The exhibited differential rate  contains, of course,  the WIMP mass dependence arising from the WIMP density in our galaxy.



One can integrate the differential rate over the recoil energy $E_R$ and multiply with the total number of nucleons to obtain the the total WIMP-Nucleus event rate as a function of the WIMP mass $m_{\chi}$. 

\section{Experimental approach for the dark matter search}

Calculations performed in previous chapters have been dedicated to the most prominent candidates, which were taken from the list given in Tab.~\ref{tab:1}. 
The DM collision with the isomeric state has an exceedingly small cross section. Therefore, the isomeric states with longer half-lives are favored due to the smaller contribution of natural decay to their total decay rates. 
The $\rm ^{180m}Ta$ is an outstanding candidate for investigation.
As mentioned above this nuclide has been proposed~\cite{PRR20} and treated by gamma-ray spectrometry with HPGe-detectors~\cite{Lehnert}. 
The expected gamma-lines of the direct isomer decay and the further $\rm ^{180}Ta$ ground state decay (half-life $8.1~\rm h$) are shown in Tab.~\ref{tab:2}. 
\begin{table}[h!]
\centering
\caption{Transition energies that can arise from the decay of the $9^-$ isomeric state of $\rm ^{180m}Ta$.}
\label{tab:2}
\begin{tabular}{ c|c|c|c|c }
 \hline
Energy (keV) & Transition & Tran. type & Tot. electron CC (Ba) & Electron energy shell (keV)\\
 \hline
37.25 &  $^{180}$Ta $9^-$ $\rightarrow$ $^{180}$Ta $2^+$ & E7 & $>$ 10$^{8}$ & L26\\
39.54 &  $^{180}$Ta $2^+$ $\rightarrow$ $^{180}$Ta $1^+$ & E1 & $>$ 0.88 & L28.5\\
76.79 &  $^{180}$Ta $9^-$ $\rightarrow$ $^{180}$Ta $1^+$ & M8 & $>$ 10$^{8}$ &K9.37; L65.8\\
93.30 &  $^{180}$Hf $2^+$ $\rightarrow$ $^{180}$Hf $0^+$ & E2 & 4.69 & K27.9; L82.8\\
103.5 &  $^{180}$W $2^+$ $\rightarrow$ $^{180}$W $0^+$ & E2 & 1.48 & K34\\
215.3 &  $^{180}$Hf $4^+$ $\rightarrow$ $^{180}$Hf $2^+$ & E2 & 0.24 & K150\\
234.0 &  $^{180}$W $4^+$ $\rightarrow$ $^{180}$W $2^+$ & E2 & 0.20 & K164.5\\
332.3 &  $^{180}$Hf $6^+$ $\rightarrow$ $^{180}$Hf $4^+$ & E2 & 0.060 & K267\\
350.9 &  $^{180}$W $6^+$ $\rightarrow$ $^{180}$W $4^+$ & E2 & 0.054 & K281,4\\
 \hline
\end{tabular}

\end{table}
The decay energies are precisely determined thanks to the recent accurate measurements of excitation energy of the isomeric state by the Penning trap mass spectrometry: $76.80(33)~\rm keV$~\cite{Nesterenko}.
The energy of transitions expected as a result of the decay of  the isomer to the ground state and the daughter nuclides of $^{180}\rm W$ and $^{180}\rm Hf$ are shown in  Tab.~\ref{tab:2}. 
Additionally, the values of the internal electron conversion coefficients indicated in the fifth  column of Tab.~\ref{tab:2}.
Only the gamma-transition with  energy $103.5~\rm keV$ has been used to search for possible response to DM~\cite{Lehnert}. The $93.3~\rm keV$ line from the electron capture is too similar to the background lines from $^{234}$Th.
The $\rm103.5~keV$ gamma-line with the branching ratio $15\%$ registered with the detector efficiency $<0.3\%$ belongs to de-excitation of daughter nuclide $^{180}\rm W$.
This results in a non-observation of any signal from the DM interaction obtained with HPGe-detector with  total photon registration efficiency of $<4\cdot10^{-4}$.

Meanwhile, Low Temperature Detectors (LTD) are widely used to search for rare events in nuclear and particle physics. 
Great success was achieved with Magnetic Micro Calorimeters (MMC)~\cite{Gastaldo, Kim} that can measure particle and photon energy with detection angle coverage close to $4\pi$.   
The use of such detectors will increase the sensitivity of recording rare events by many orders of magnitude in comparison with the conventional germanium detector used in~\cite{Lehnert}.
In addition to the angular advantages in such detectors, it is possible to efficiently register the transition energies transmitted by internal conversion electrons, which considerably prevail in the decays of isomeric states with low transition energies.
Another very important advantage is the energy resolution, which exceeds the extreme limits of any semiconductor detectors. 
The MMC-cryogenic detector consists of a metallic absorber that stores the released energy and has a very good connection to the temperature sensor -- a paramagnetic alloy, which resides in a low magnetic field. 
The sensor is weakly connected to a thermal bath.
The energy released in the absorber from a radioactive source completely enclosed in the absorber raises the temperature of the detector.
This leads to a change of the sensor magnetization which is read out as a change of flux by SQUID magnetometer. 
Such type of an MMC  has been used in the spectra measurement of $^{163}\rm Ho$ within the ECHo project~\cite{Gastaldo}. 
The energy resolution on the level of $1.2~\rm eV$ was achieved at X-ray energy of $6~\rm keV$. 
The same detector can be used to search for DM-particles with  interaction to isomeric state of  $\rm ^{166m}Ho$ (see Tab.~\ref{tab:1}). 
This isotope can plentifully be produced in reactors by neutron irradiation of $^{165}\rm Ho$.
However, the small cross section for the DM-scattering obtained in our calculations for $\rm ^{166m}Ho$ makes this detection unrealistic due to the strong background contribution of events from natural $\rm ^{166m}Ho$ radioactive decay. 

We emphasize again that $\rm ^{180m}Ta$ is a promising candidate because its extremely long lifetime has very small background from its natural radioactive decay. 
Unlike the germanium detector, the MMC-technique allows measurement of all decay channels: characteristic X-ray and gamma-radiation and equally importantly, peaks from internal conversion electrons, which prevail in the spectrum of low-energy transitions, as can be seen in the fourth column of Tab.~\ref{tab:2}. 
The last column of the same table shows intensive expected electron energies of these transitions. 
The most intensive electron spectrum belongs to the energy region below $100~\rm keV$.
The most prominent characteristic K X-rays for Ta and Hf nuclides that follow the electron conversion process in the interval from $64.4$ to $67.0~\rm keV$ are beyond the electron conversion spectra and can be also detected by the MMC-detector with high precision of approximately $100~\rm eV$. 
Thus, in comparison with the germanium detector used in the paper~\cite{Lehnert}, the MMC-method can provide plenty of indicators of DM interactions with the Ta-isomeric state.
As a matter of fact some careful elaboration concerning the long-term stability in vacuum and maintenance at low temperature, as well as the physicochemical properties of Ta can be taken into account. 

The $\rm^{180m}Ta$ isomeric state is the only one in nature, having the abundance of $0.012\%$ in natural tantalum. 
Isotopic extraction of tantalum is very difficult. However $3~\rm mg$ of isomer diluted in $30~\rm mg$ of $\rm^{181}Ta$ was already used in~\cite{laser_phys}. 
Hopefully, a similar target is achieved by our proposed experiment instead of the method of~\cite{Lehnert} in which $180~\rm mg$ of $\rm^{180m}Ta$ in $1.5~\rm kg$ of natural tantalum is used.
The total achievable mass of $3~\rm mg$ for $\rm^{180m}Ta$ corresponds to $N_N \approx 10^{19}$ and this value is later used for the half-life calculation in~\ref{sec:half_life}. 

Another possibility to produce $\rm^{180m}Ta$ is the reaction $(p,pn)$ on $\rm^{181}Ta$. 
Readout system can use SQUID microwave multiplexing~\cite{Gastaldo}.
The natural background will be approximately 1 event per year if the expected total lifetime of this isomer is $10^{19}$ year~\cite{Lehnert_1}.
As can be seen from these estimates, the observation of the DM-effect is very challenging.
It becomes more realistic if the cross section attributable to DM de-excitation and/or the natural lifetime of the isomeric state deviate for two-three orders of magnitude from used one.
Such deviations are entirely possible.
As long as no signature is observed in the long term experiment, limits can be constrain the DM de-excitation. 
However, if the $\rm^{180m}Ta$ decay is observed, then measurements with increased exposure time in deeper underground locations may be indicative of a DM response.

\section{Half-life time estimation}
\label{sec:half_life}

The $\rm^{180}Ta$ isomer is predicted to decay to the ground state, but the half-life has not yet been determined.
There are only experimental and theoretical upper limits available for this isotope, since decay has never been observed in a real experiment.
The expected half-life time should exceed $10^{19}$ years~\cite{Lehnert_1}.
And the lower experimental bound for the total half-life time is $4.5\cdot10^{16}$ years (90\% C.L.)~\cite{Lehnert_2017}.
The interaction with the WIMP and subsequent decay can be interpreted as normal decay, and from this the half-life time can be estimated.

As shown in the previous section, the overall background is one event per year.
Since all DM searches do not observe any events (signal is zero),  statistical evaluation can be simplified using the conventional Feldman-Cousins approach~\cite{Feldman:1997qc}.
The expected number of signal events $n_{exp}$ can be written as
\begin{equation}
\label{eq_1}
    n_{exp} = \Phi \cdot \sigma_{\rm tot} \cdot T \cdot N_N \cdot \varepsilon,
\end{equation}
where $\Phi$ WIMP flux, $\sigma_{\rm tot}$ total cross-section of WIMP isomer interaction, $T$ exposure time, $N_N$ number of isomer atoms, $\varepsilon$ detection efficiency.
With zero signal  this $n_{exp}$ can be associated with a particular upper limit $n_{up}$. 
For instance in our case the 95\% C.L. $n_{up}$ equals 2.33 (signal zero, background one for 1 year exposure~\cite{Feldman:1997qc}).

From another side, signal can be interpreted as normal radioactive decay that follows the exponential decay law. In this case $n_{exp}$ can be expressed as
\begin{equation}
\label{eq_1a}
n_{exp} = \varepsilon N_N \Big(1 - \exp{\Big[-\ln{2} \cdot \frac{T} {T_{1/2}}\Big]}\Big) \simeq \frac{\varepsilon N_N T}{T_{1/2}} \cdot \ln{2},
\end{equation}
with expected half-life time $T_{1/2}$.
Combining Eq.~\eqref{eq_1} and Eq.~\eqref{eq_1a}, we can express $T_{1/2}$ as a function of $T$
\begin{equation}
\label{eq_2}
T_{1/2} = \frac{T \cdot \ln{0.5}}{\ln{(1 - \Phi \sigma T)}} \simeq \frac{\ln{2}}{\Phi \sigma}
\end{equation}
As can be seen, the $T_{1/2}$ does not depend on the amount of isomer and detection efficiency,  because in our case the cross-section is already determined from theory.
Based on that, we can estimate the sensitivity of the proposed experimental setup to the WIMP-nucleon interaction, where the last one can be interpreted as normal radioactive decay.  
\begin{figure}
    \centering
    \includegraphics[width=0.7\textwidth]{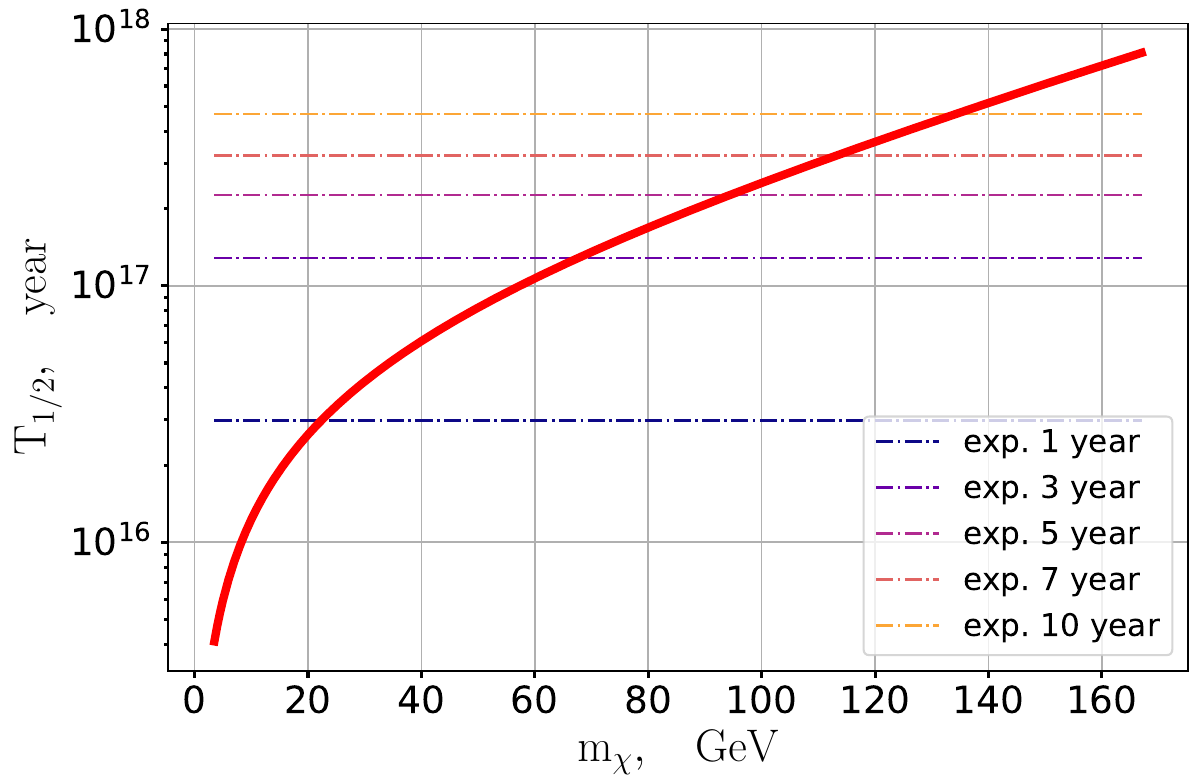}
    \caption{The solid red line is the half-life time as a function of the WIMP mass for $\rm ^{180m}Ta$. 
    Dashed lines indicate 95\% C.L. sensitivity for the proposed experimental setup with different exposures. 
    Assumed parameters for calculating sensitivity $\varepsilon = 0.01$, $N_N = 10^{19}$.}
    \label{fig:T_12}
\end{figure}
In Fig.~\ref{fig:T_12} the half-life time of  $\rm ^{180m}Ta$ as function of WIMP mass is depicted.
Several levels of the experimental sensitivity with different exposure time are also shown.
We can conclude that the WIMP-nucleon interaction can be measured with 95\% C.L. for masses $\leq 130$~GeV (10 years exposure time is assumed). 
The conservative value of the detector efficiency was chosen as 1\%, and the result depends linearly from $\varepsilon$, hence it can be easily scaled for higher values of $\varepsilon$. 

\section{Discussion}

We have seen  that, not  unexpectedly,   the nuclear ME encountered in the inelastic WIMP-nucleus scattering involving isomeric nuclei is much smaller than  that involved in the elastic process considered in the standard WIMP searches. This occurs  for two reasons: a) the form factor in the elastic being  favorable, see Fig. \ref{fig:FFelHo}, and b) in the elastic case the cross section is proportional to the mass number  $A^2$.  In the present case the NME for $\rm^{180}Ta$, as indicated by the coefficients  appearing in  Eq. (\ref{Eq:coefNilsson}), is  not unusually small compared to other typical  inelastic processes. The Nilsson model is expected to work well in the case  of $\rm^{180}Ta$, but the obtained event rate is quite small. 
At the same time, our calculations demonstrated a significant suppression in the matrix element for  $\rm ^{166}Ho$, that allowed us to rule out it from the list of candidates for experimental proposal.

Following the estimated WIMP-nucleon cross section and  the current mass of $\rm ^{180}Ta$ isomer ($N_N\approx10^{19}$), the estimated half-life time is between $10^{15}$ and $10^{18}$ years, covering current experimental limit $4.5\cdot10^{16}$ years (90\% C.L.)~\cite{Lehnert_2017}. 
However it is still not enough to reach the expected half-life time for $\rm ^{180m}Ta$. 
That is why various experimental approaches are required to disentangle this puzzle.
Further improvement can be achieved using larger mass of isomer with better detection efficiency.
At the same time, the experiment can exploit the signal provided by the subsequent standard decay of the $2^+$ state to the ground state, that is not available in the conventional WIMP searches. 
 
 \appendix
 
\section{Recoil energy calculation}
\label{app:rec_energy}

Let us estimate the maximum recoil energy. From Eq. (\ref{Eq:maxq1}) we find the maximum momentum is given by:
\begin{equation}
\frac{1}{q_{\rm max}}\Big(\frac{q_{\rm max}^2}{2 \mu_r}-\Delta\Big)=\upsilon_{\rm esc}
\end{equation}
The only acceptable solution is
\begin{equation}
q_{\rm max}=\mu_r \upsilon_{\rm esc}\left [1+\sqrt{1+\frac{2 \Delta} {\mu_r \upsilon^2_{\rm esc}}}\right]
\end{equation}
Thus the maximum recoil energy is given by
\begin{equation}
\left(E_R\right)_{\rm max}=\frac{1}{2 m_A} q_{\rm max}^2=\frac{1}{2}m_A\upsilon_{\rm esc}^2\frac{1}{(1+x)^2}\left [1+\sqrt{1+\frac{\Delta(1+x)}{\frac{1}{2}m_A\upsilon_{\rm esc}^2}}\right]^2,\quad x=\frac{m_A}{m_\chi}
\end{equation}

In the special case of elastic scattering ($\Delta=0$), we find the expected results
\begin{equation}
\left(E_R\right)_{\rm max}=2m_A\upsilon_{\rm esc}^2, ~ m_{\chi}>>m_A, ~ \left(E_R\right)_{\rm max}\approx 0,~ m_A>>m_{\chi}, ~ \left(E_R\right)_{\rm max}=\frac{1}{2}m_A\upsilon_{\rm esc}^2, ~ m_A\approx m_{\chi}
\label{Eq:MaxER}
\end{equation}
In the case of $\Delta>0$  we find that for $m_{\chi}\approx m_A$
\begin{equation}
\left(E_R\right)_{\rm max}=\frac{1}{8}m_A\upsilon_{\rm esc}^2\left [1+\sqrt{1+\frac{2 \Delta}{\frac{1}{2}m_A\upsilon_{\rm esc}^2}}\right]^2\approx0.5\cdot10^{-6 }m_A \left [1+\sqrt{1+\frac{ \Delta}{2\cdot10^{-6}m_A}}\right]^2,\, m_A\approx m_{\chi}
\end{equation}

\section{The differential  cross section  for the inelastic WIMP-nucleus scattering}
\label{app:diff_xsec}

Eq.\eqref{Eq.DifCros}, using Eq.\eqref{Eq.TotCrosN2}, can  be cast in the form:
\begin{equation}
d\sigma=\Lambda \frac{\sigma_N}{m_N^2}  \frac{1}{\upsilon}\frac{1}{(2 \pi)^2}d^3 {\bf q}\delta\left (\frac{q^2}{2 \mu_r}-q \upsilon \xi-\Delta\right )\frac{|ME(q^2)|^2|}{f_V^2+3 f_A^2}
\label{Eq.DifCros2}
\end{equation}
Where 
$$\Lambda=\frac{2 \pi}{4}.$$
 Folding Eq.\eqref{Eq.DifCros2} with the velocity distribution we find\footnote{The factor $1/\upsilon_0$, with dimension of inverse velocity, was introduced for convenience. A compensating  factor   $\upsilon_0$ will be  used in  multiplying the particle density obtaining the flux. Thus we get the traditional formulas,  flux=particle density $\times$ velocity and rate= flux $\times$ cross section.}
 \begin{equation}
 \begin{split}
\frac{1}{\upsilon_0}\frac{1}{\sigma_N}\langle \upsilon  \frac{d\sigma}{dE_R}\rangle =&\Lambda \frac{m_A}{m_N^2} \frac{1}{\upsilon_0}\frac{1}{2 \pi}   \frac{|ME(q^2)|^2}{f_V^2+3 f_A^2} \\
&\left[ \left (\Theta\left (\Delta-\frac{M_AE_R}{\mu_r}\right ) \right )\int_{\upsilon_1}^{\upsilon_{\rm esc}}K(\upsilon)d \upsilon + \left (\Theta\left(-\Delta+\frac{M_AE_R}{\mu_r}\right ) \right )\int_{\upsilon_2}^{\upsilon_{\rm esc}}K(\upsilon)d \upsilon \right ]
\end{split}
\end{equation}
where $ E_R$ is the nuclear recoil energy and  $K(\upsilon)$ given by the velocity distribution
 \begin{equation}
K(\upsilon)=\int d \Omega(\hat{\upsilon}) \upsilon f_{\rm distr}(\bm{v})
\end{equation}
and $\Theta$ is the step function:
 \begin{equation}
\Theta(x)=\left \{\begin{array}{cc}1,&x>0\\0,&x<0\\
	\end{array}\right .
\end{equation}
Furthermore
 \begin{equation}
\upsilon_{1,2}=\pm~\frac{1}{q}\Big(\Delta-\frac{q^2}{2\mu_r}\Big)
\label{Eq.vminofq}
\end{equation}
Note the dependence of the cross section on the recoil energy comes in two ways: i) From the nuclear form factor and ii) from the minimum required velocities $\upsilon_1$ and $\upsilon_2$ in the folding with the velocity distribution.

We will specialize our results in the commonly used Maxwell- Boltzmann (MB) distribution in the local frame
 \begin{equation}
f_{MB}=\frac{1}{\pi^{3/2}}\frac{1}{\upsilon^3_0}e^{-(y^2+2 y \xi+1)}, \quad y=\frac{\upsilon}{\upsilon_0}
\label{Eq:M_B}
\end{equation}
where $\upsilon_0$ is the velocity of the sun around the center of the galaxy and $\xi$ the angle between $\bm{v}$ and the direction of sun's motion. Then
 \begin{equation}
K(\upsilon)=\frac{2}{\sqrt{\pi}}e^{-(y^2+1)}y\int_{-1}^1e^{-2 y \xi}=\frac{2}{\sqrt{\pi}}e^{-(y^2+1)}\sinh{2 y}
\end{equation}
 \begin{equation}
 \begin{split}
&\frac{1}{\upsilon_0}\frac{1}{\sigma_N}\langle \upsilon  \frac{d\sigma>}{dE_R}\rangle =\Lambda \frac{m_A}{m_N^2}\frac{1}{\upsilon^2_0}\frac{1}{2 \pi}   \frac{|ME(q^2)|^2|}{f_V^2+3 f_A^2} \\
&\left[ \left (\Theta\left (\Delta-\frac{M_AE_R}{\mu_r}\right ) \right )\int_{y_1}^{y_{\rm esc}}e^{-(y^2+1)}\sinh{2y}dy + \left (\Theta\left(-\Delta+\frac{M_AE_R}{\mu_r}\right ) \right )\int_{y_2}^{y_{\rm esc}}e^{-(y^2+1)}\sinh{2y}dy \right ]
\end{split}
\end{equation}
The above integrals can by computed analytically
\begin{equation}
\begin{split}
\frac{1}{\upsilon_0}\frac{1}{\sigma_N}\langle \upsilon  \frac{d\sigma}{dE_R}\rangle =&\Lambda \frac{m_A}{m_N^2}\frac{1}{\upsilon^2_0}\frac{1}{2 \pi}  \frac{|ME(q^2)|^2|}{f_V^2+3 f_A^2} \\
&\left[ \left (\Theta\left (\Delta-\frac{M_AE_R}{\mu_r}\right ) \right )\psi_1(y_1,y_{\rm esc})+ \left (\Theta\left(-\Delta+\frac{M_AE_R}{\mu_r}\right ) \right )\psi_2(y_2,y_{\rm esc})\right ]
\end{split}
\end{equation}
where
\begin{equation}
\begin{split}
\psi_i(y_i,y_{\rm esc})&=\frac{1}{4} \sqrt{\pi }
\left(\text{erf}\left(1-y_i\right)+\text{erf}\left(y_i+1\right)\right)-\frac{1}{4}\sqrt{\pi }\left(\text{erf}\left(1-y_{\rm esc}\right)+\text{erf}\left(y_{\rm esc}+1\right)\right), \quad i=1, 2
\label{Eq:yoferf}
\end{split}
\end{equation}
The functions $\psi_i(y_i,y_{\rm esc})$ depend on the momentum transfer. This depends on the specific nuclear target and will be discussed below.
The expression contained in the last square bracket is momentum dependent  and provides a restriction in the range and distribution of momentum.
It is  given by Figs.~\ref{fig:psi(q)Ta} and~\ref{fig:psi(q)Ho}  for the nuclei of interest in this work.

Sometimes is useful to modify the above formulas using dimensionless variables. let us define $\eta=q R$, where is the nuclear radius. Then
$$ \frac{1}{q}\Big(\frac{ q^2}{2 \mu_r}-\Delta\Big)=\left (\frac{1}{2}(1+x)\frac{\eta^2}{m_A R^2}-\Delta \right ) \frac{R}{\eta}=\frac{1}{2}(1+x)\frac{\eta}{a}-\frac{b}{\eta}, ~ a=m_A R , ~ b=\Delta R$$
Thus
\begin{equation}
y_1=\frac{c}{\upsilon_0}\left (-\frac{1}{2}(1+x)\frac{\eta}{a}+\frac{b}{\eta}\right ),\,y_2=\frac{c}{\upsilon_0}\left (\frac{1}{2}(1+x)\frac{\eta}{a}-\frac{b}{\eta}\right )
\label{Eq:yofu}
\end{equation}
\begin{equation}
\begin{split}
\frac{1}{\upsilon_0}\frac{1}{\sigma_N}\langle \upsilon  \frac{d\sigma>}{dE_R}\rangle =&\Lambda \frac{m_A}{m_N^2}\frac{1}{\upsilon^2_0}\frac{1}{2 \pi}   \frac{|ME(q^2)|^2|}{f_V^2+3 f_A^2} \\
&\left[ \left (\Theta\left (b-\frac{1}{2}(1+x)\frac{\eta^2}{a}\right ) \right )\psi_1(y_1,y_{\rm esc})+ \left (\Theta\left(-b+\frac{1}{2}(1+x)\frac{\eta^2}{b}\right ) \right )\psi_2(y_2,y_{\rm esc})\right ]
\label{Eq:finalexpr}
\end{split}
\end{equation}
where the functions $\psi_1$ and $\psi_2$ are given by Eq.\eqref{Eq:yoferf} via Eq.\eqref{Eq:yofu}.

In shell model calculations instead of the nuclear radius $R$  one may use the harmonic oscillator  size parameter $b_A$ (see Appendix~\ref{app:shell_model}).
There remains the crucial part of the calculation in involving the NME and the associated nuclear form factor. 

\section{The nucleon cross section and event rate}
\label{app:NWimpRate}

\begin{figure}
\centering
\includegraphics[width=0.7\textwidth]{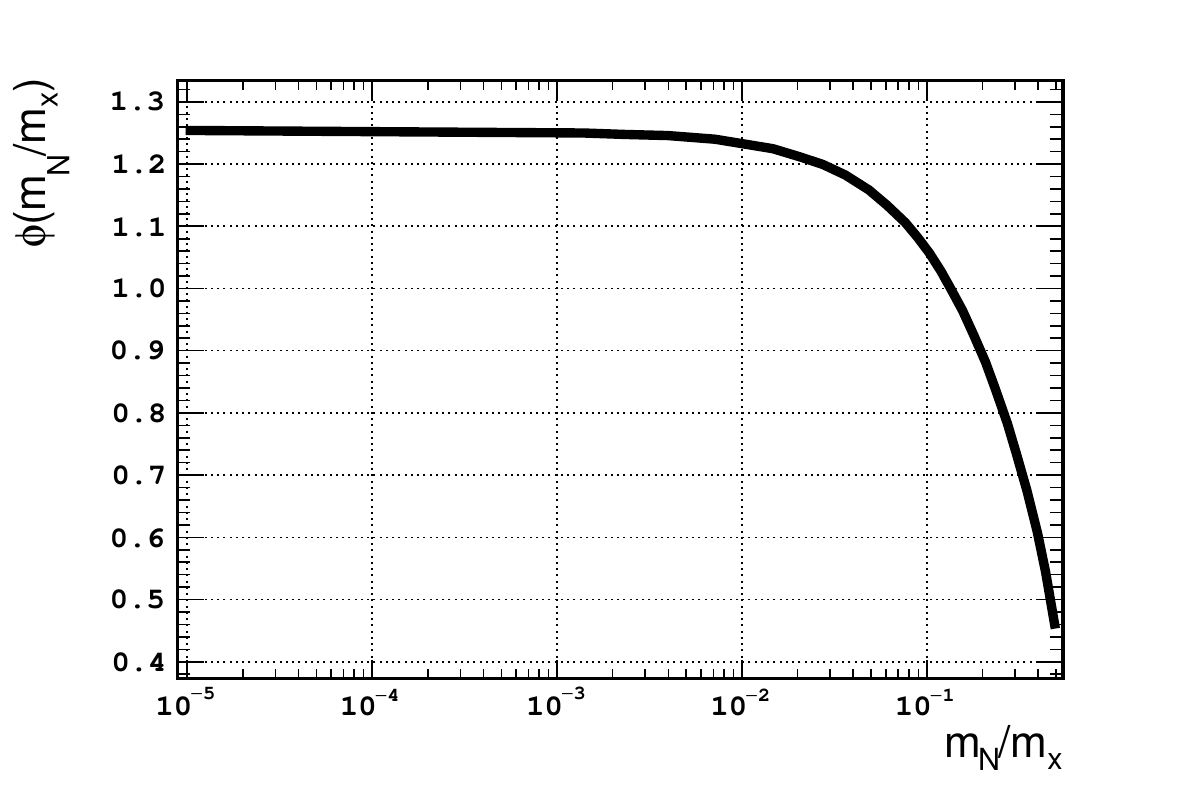}
\caption{The dependence on the function $\phi(x)$ on the mass $m_{\chi}$ appearing  in the case of the nucleon-WIMP scattering}
\label{sigmaNxa}
\end{figure}

We have seen that Eq.\eqref{Eq.DifCrosN} is correct, but it does not indicate the range of $q$ involved. 
To find it we return to the basic expression $\frac{q^2}{2 mu_2}-q \upsilon \xi=0$.
This leads to 
\begin{equation}
\xi=\frac{q}{2 \mu_2}{ \upsilon}<1\Rightarrow \upsilon>\upsilon_{min},\quad\upsilon_{min}=\frac{\sqrt{2 m_N E_R}}{2 m_N}(1+x)=\sqrt{\frac{E_R}{2 m_N}} (1+x)
\end{equation}
This implies that 
\begin{equation}
y_{min}=\sqrt{\frac{E_R}{2 m_N\upsilon_0^2}} (1+x)
\end{equation}
Folding expression of Eq.\eqref{Eq.DifCrosN} with the velocity distribution, we obtain
\begin{equation}
\langle \upsilon d\sigma_N(\upsilon)\rangle=  \int_{\upsilon_{min}}^{\upsilon_{\rm esc}}\upsilon d\sigma_N(\upsilon)f_{dist}(\bm{v}) d^3 \bm{v}
\label{Eq.TotCrosN0a}
\end{equation}
For a Maxwell-Boltzmann distribution,  see Eq.\eqref{Eq:M_B}, one can show that :
\begin{equation}
\langle d \sigma_N(\upsilon)\rangle =Sp \,dE_R\, \upsilon_0 \frac{2}{\sqrt \pi}\int_{y_{min}}^{y_{\rm esc}}y f_{MB}dy,\,Sp=\frac{1}{\upsilon^2_0}\frac{1}{2 \pi} m_N  \left (\frac{G_F}{\sqrt{2}}\right )^2|ME_N|^2
\label{Eq.TotCrosN0b}
\end{equation}
where the  factors $\upsilon_0$ have been separated judiciously (in DM searches the factor $\upsilon_0$ is absorbed in the WIMP flux (rate=flux $\times$ cross section).\\
Performing the integration over $y$ we find 
\begin{equation}
\langle d \sigma_N(\upsilon)\rangle = \upsilon_0 \psi(y_{min},y_{\rm esc})Sp.
\end{equation}
with 
\begin{equation}
\psi(y_{min},y_{\rm esc})=\frac{1}{4} \sqrt{\pi }
\left(-\text{erf}\left(1-y_{\text{\rm esc}}\right)-\text{erf}\left(y_{\text{\rm esc}}+1\right)+\text{erf}\left(1-y_{min}\right)+\text{erf}\left(y_{\min }+1\right)\right)
\end{equation}
$$y_{min}=\sqrt{\frac{E_R}{2 m_N\upsilon_0^2}}(1+x) =\sqrt{\frac{E_R}{\mbox{keV}}}(1+x)$$ 
We must  now integrate over the recoil energy from zero to $(E_R)_{\rm max}$ given by Eq.\eqref{ERMax}
\begin{equation}
\langle  \sigma_N(\upsilon)\rangle = \upsilon_0 (4 m_N \upsilon^2_0 )\,Sp \phi(x),\quad\phi(x)=\int_0^{(E_R)_{\rm max}}dE_{R} \psi(y_{min},y_{\rm esc}).
\end{equation}
 The function can only be obtained numerically. It is exhibited in Fig.~\ref{sigmaNxa}. It is clear that for  values of $m_{\chi}$ much larger than $m_N$ the cross section becomes independent of $m_{\chi}$ .
 
Adopting the value $\phi(x)  \approx 1$ we obtain:
\begin{equation}
\langle  \sigma_N(\upsilon)\rangle = \upsilon_0 \,4.0\, \frac{1}{2 \pi} m^2_N  \left (\frac{G_F}{\sqrt{2}}\right )^2|ME_N|^2
\end{equation}
The WIMP-nucleon interaction is not known. Let us assume  that it is of $V-A$ type  in dimensionless units. Then
\begin{equation}
|ME_N|^2=|ME_F|^2+|ME_A|^2=f_V^2+f_A^2\langle N|\bm{s}|N \rangle ^2=f_V^2+3f_A^2
\end{equation}
Thus, if we absorb the factor $\upsilon_0 $ into the flux, we can write the effective total cross section becomes:
\begin{equation}
\sigma_N=4.0 \frac{1}{2 \pi} m^2_N  \left (\frac{G_F}{\sqrt{2}}\right )^2  \left (f_V^2+3 f_A^2\right )
\label{Eq.TotCrosN}
\end{equation}
$\sigma_N=3,3 \times 10^{-39}(f_V^2+3 f_A^2)$\\
For calculating the rates in addition to the cross section one needs the flux of the oncoming particles and the number of particles in the target. For an estimate we  will consider $N_N=10^{24}$ particles in the target.
The  WIMP energy density in our vicinity is $\rho=0.3$~$\rm GeV\cdot cm^{-3}$, leading to a particle density $N_{\chi}=\frac{\rho}{m_{\chi}}=\frac{\rho}{m_{N}}x$ with $x=\frac{m_N}{m_{\chi}}$\\
The flux is the particle density times the WIMP velocity:
\begin{equation}
\label{Eq.Flux}
\Phi=N_{\chi} \upsilon=\frac{\rho}{m_{\chi}}\upsilon_0=\frac{\rho}{m_{N}}\upsilon_0 x=(0.3 \times (220 \times 10^3\times 10^2)  \times 3.157\times 10^7y^{-1}x=2.1 \times 10^{14}\mbox{cm}^{-2}y^{-1} x
\end{equation}
The combined effect on the rate is 
\begin{equation}
\label{Eq.Comb_rate}
f_R=\Phi N_N \sigma_N=2.1 \times 10^{38}\mbox{cm}^{-2}y^{-1} x\times  3.3 \times 10^{-39}(f_V^2+3 f_A^2)
\end{equation}
Thus
$$f_R =0.72 y ^{-1} x (f_V^2+3 f_A^2)$$
for $m_{\chi}=m_N$ as reference we obtain the rate:
\begin{equation}
\label{Eq.rate_appendix}
R=0.72 \mbox{y}^{-1}  (f_V^2+3 f_A^2)
\end{equation}
A similar procedure can applied in the WIMP-nucleus-scattering with the obvious modification $m_N\rightarrow m_A$.

\section{The Reduced nuclear ME for a two particle system}
\label{app:reduced_ME}

As we have seen, the two particle proton- neutron system of the type considered here can be described in terms of one proton and one neutron Nilsson orbitals. These can be expanded in a spherical basis. The reduced nuclear matrix element (RNME) of WIMP-nuclear transition can be obtained using the well known Racah techniques.\\
i) {\bf  The RNME in the case of $\rm^{180}Ta$ target.} \\
Since the initial state is of negative parity involving a proton in a negative parity state, while the final state considered here is of positive parity, the interaction involves only the proton component with neutron being a spectator particle. Thus the reduced nuclear matrix takes the form:
\begin{equation}
\langle [j'_1j_2]J_f||T^{[\lambda \otimes s]J}||[j_1j_2]J_i\rangle=\sqrt{\frac{2J_f+1}{2j'_1+1}}U[j_1,j_2,J_i,J,0,J,j'_1j_2,J_]]\langle j'_1||T^{[\lambda \otimes s]}||j_1\rangle
\end{equation}
where $U[\cdots]$ is the unitary Nine-J symbol, see e.g. \cite{Hecht2000}. Furthermore:
\begin{equation}
\langle j'_1||T^{[\lambda \otimes s]J}||j_1\rangle=\sqrt{2j'_1+1} U[\ell_1,1/2,j_1,\lambda, s ,J,\ell'_1,1/2,j'_1]	\frac{\langle \ell'_1||\sqrt{4 \pi+1}Y^{\lambda}||\ell_1\rangle}{\sqrt{ 2 \ell'_1+1}}\times\begin{array}{cc}f_V,s=0\\f_A \sqrt{3}, s=1\\ \end{array}
\end{equation}
The last reduced ME is essentially that of the spherical harmonic
\begin{equation}
\langle \ell'_1||\sqrt{4 \pi+1}Y^{\lambda}||\ell_1\rangle=\sqrt{(2 \ell_1+1)(2 \lambda+1)} \langle \ell_1,0;\lambda, 0|\ell'_1, 0\rangle\langle n'_1\ell'_1|j_{\lambda}( k r )|n_1\ell_1\rangle \Rightarrow
\end{equation}
\begin{equation}
\begin{split}
&\langle [j'_1j'_2]J_f||T^{[\lambda \otimes s]J}||[j_1j_2]J_i\rangle= \\
&\delta_{j_2,j'_2}\sqrt{2J_f+1}U[j_1,j_2,J_i,J,0,J,j'_1j_2,J_f] U[\ell_1,1/2,j_1,\lambda, s ,J,\ell'_1,1/2,j'_1]
\frac{\sqrt{(2 \ell_1+1)(2 \lambda+1)}}{\sqrt{(2 \ell'_1+1)}}\times \\
&\langle \ell_1,0;\lambda, 0|\ell'_1, 0\rangle\langle n'_1\ell'_1|j_{\lambda}( k r )|n_1\ell_1\rangle
\label{Eq:PartialRNME}
\end{split}
\end{equation}
ii){\bf The case of interacting protons and neutrons.}\\
In case that both protons and neutron can interact the above expression can be written compactly as follows
\begin{equation}
\begin{split}
&\langle [j'_1j'_2]J_f||T^{[\lambda \otimes s]J}||[j_1j_2]J_i\rangle= \\
&\sqrt{2J_f+1}\left (\delta_{j_2,j'_2}U[j_1,j_2,J_i,J,0,J,j'_1j'_2,J_f]ff(j_1,j'_1)+ \delta_{j_1,j'_1}U[j_1,j_2,J_i,0,J,J,j'_1j'_2,J_f]ff(j_2,j'_2 )\right )
\label{Eq:FullRNME}
\end{split}
\end{equation}
with
\begin{equation}
ff(j_i,j'_i)=\frac{\sqrt{(2 \ell_i+1)(2 \lambda+1)}}{\sqrt{(2 \ell'_i+1)}}U[\ell_i,1/2,j_i,\lambda, s ,J,\ell'_i,1/2,j'_i]
\langle \ell_i,0;\lambda, 0|\ell'_i, 0\rangle\langle n'_i\ell'_i|j_{\lambda}( k r )|n_i\ell_i\rangle,\quad i=1,2
\end{equation}
A detailed explicit calculation reduced matrix elements in the case of $^{166}$H$\mbox{o}$ is given in section~\ref{sec:pnHo}.

\subsection{The explicit calculation reduced matrix elements in the case of $\rm^{166}Ho$} 
\label{sec:pnHo}

Using standard  Racah techniques one finds \cite{Hecht2000}:
\begin{equation}
\begin{split}
RME=&C^2_{0h_{11/2}}C^2_{0g_{11/2}}\left [ \begin{array} {ccc}1_1/2&11/2&0 \\ 7& 0& 7\\ 11/2&11/2 &7\\ \end{array} \right ] \\
& \frac  {\langle 0g11/2||^{T(\lambda,J)}||0g{11/2}\rangle}{\sqrt{12}}+\frac{C_{0g_{13/2}}}{C^2_{0g_{11/2}}}
\frac  {\langle0g11/2||T^{(\lambda,J)}||0g_{13/2}\rangle}{\sqrt{12}}\sqrt{\frac{7}{6}}- \frac {\langle 0h11/2||T^{(\lambda,J)}||0h_{11/2}\rangle}{\sqrt{12}}                
\end{split}
\end{equation}
the unitary nine-j being $1/\sqrt{15}$.
$$\frac  {\langle 0g11/2||^{T(\lambda,J)}||0g{11/2}\rangle}{\sqrt{12}}=\left [ \begin{array} {ccc}6&1/2&11/2 \\ \lambda& 1& J\\ 6&1/2 &11/2\\ \end{array} \right ] \frac{\langle 6||\sqrt{4 \pi} 
	Y^{\lambda}||6\rangle}{\sqrt{13}}\frac{\langle 1/2|||\bm{\sigma}||1/2\rangle}{\sqrt{3}}=$$ $$\left [ \begin{array} {ccc}6&1/2&11/2 \\ 
\lambda& 1& J\\ 6&1/2 &11/2\\ \end{array} \right ]\sqrt{2 \lambda+1},\langle (6,0),(\lambda,0)(6.0)\rangle \sqrt{3}
=\begin{array}{cc}140 \sqrt{\frac{5}{46189}},&\lambda=8\\
\frac{10\sqrt{\frac{70}{209}}}{13},&\lambda=6\\
\end{array}
$$
with the unitary nine-j being $2 \sqrt{\frac{70}{263}}$ and $-\frac{1}{13}\sqrt{\frac{35}{6}}$ for $\lambda$=8 and 6 respectively.\\
The expression  for the $\frac  {\langle 0g11/2||^{T(\lambda,J)}||0g{13/2}\rangle}{\sqrt{12}} $ is similar except that now the  nine-j are $ -\sqrt{\frac{35}{442}}$ and $-\frac{4}{13}\sqrt{\frac{10}{7}}$ for $\lambda=8$ and 6 respectively.\\
Furthermore	
$$
\frac  {\langle 0h11/2||^{T(\lambda,J)}||0h{11/2}\rangle}{\sqrt{12}}=\left [ \begin{array} {ccc}5&1/2&11/2 \\ \lambda& 1& J\\ 5&1/2 &11/2\\ \end{array} \right ] \frac{\langle 5||\sqrt{4 \pi} Y^{\lambda}||5\rangle}{\sqrt{11}}\frac{\langle 1/2|||\bm{\sigma}||1/2\rangle}{\sqrt{3}}=
$$
$$
\left [ \begin{array} {ccc}5&1/2&11/2 \\ \lambda& 1& J\\ 5&1/2 &11/2\\ \end{array} \right ]\sqrt{2 \lambda+1},\langle (5,0),(\lambda,0)(5.0)\rangle \sqrt{3}
\begin{array}{cc}
-28 \sqrt{\frac{5}{46189}},&\lambda=8\\
-\frac{2
	\sqrt{\frac{1330}{11}}}{13},&\lambda=6\\
\end{array}
$$
now the relevant  nine-j are $ -2\sqrt{\frac{2}{561}}$ and $\sqrt{\frac{95}{858}}$ for $\lambda$=8 and 6 respectively.

\subsection{The explicit calculation reduced matrix elements in the case of $^{180}$T$\mbox{a}$}

i) Vector matrix element.\\
$$
 VME=\frac{f_V}{f_A}(1.52946 F_{6,5,7}(u)+1.79799
 F_{6,5,9}(u)+2.19718
 F_{6,5,11}(u)+1.52946
 F_{6,3,7}(u)+$$ $$1.52946 
 F_{6,3,9}(u),1.01419+
 F_{4,5,7}(u)+1.01419
 F_{4,5,9}(u)-0.06445
 F_{4,3,7}(u))
 $$
 
 i) Axial vector matrix element.\\
 $$
 AME=-2.04512
 F_{6,5,7}-2.31181
 F_{6,5,9}(u)-3.58938
 F_{6,5,11}(u)-2.04512
 F_{6,3,7}(u)$$ $$-3.3217
 F_{6,3,9}(u)-2.05117
 F_{4,5,7}(u)-2.16715
 F_{4,5,9}(u)-0.3215
 F_{4,3,7}(u)
 $$
 
\section{The shell model nuclear form factors}
\label{app:shell_model}

The needed form factors $FF_{\ell\lambda}(u)$ are as follows\\
	i) {\bf The case of $\rm^{166}Ho$.}\\
	$$
	FF_{6,8}(u)=\frac{e^{-\frac{u^2}{4}} u^8
		\left(u^4-84
		u^2+1596\right)}{8648640},\,
	FF_{6,6}(u)=-\frac{e^{-\frac{u^2}{4}} u^6
		\left(u^6-114 u^4+3876
		u^2-38760\right)}{8648640},$$ $$
	FF_{5,8}(u)=-\frac{e^{-\frac{u^2}{4}} u^8
		\left(u^2-38\right)}{332640},\,
	FF_{5,6}(u)\frac{e^{-\frac{u^2}{4}} u^6
		\left(u^4-68
		u^2+1020\right)}{332640}$$
	with $u=b_N(A) q$ with $b_N(A)$, the harmonic oscillator size parameter given, by:
\begin{equation}
b_N(A)\approx 1.00 \sqrt[6]{A}
\label{Eq:NucBN}
\end{equation}
The first index specifies the interaction orbit (the transition is diagonal) and the second index corresponds to the multipolarity $\lambda$.
	
ii) {\bf The case of $\rm ^{180}Ta$.}\\
	One can also  calculate the odd parity form factors analytically. The resulting expressions are rather  complicated to present here. We are satisfied with exhibiting  our results in Fig. \ref{fig:SMFFTa}.
	
iii) In the case of the Helm type form factors one uses the nuclear radius	
\begin{equation}
	R(A)=1.24\cdot A^{1/3}
	\label{Eq:NucRad}
\end{equation}

\section{Some useful expansions of Nilsson levels to shell model states}
\label{app:nilss_levels}

\begin{table}[htb]
\centering
\caption{Expansions of neutron Nilsson orbitals $\Omega[N n_z \Lambda]$ in the shell model basis $|N l j \Omega \rangle$ for three different values of the deformation $\epsilon$. 
	}
\label{tab:3}       
	
	\begin{tabular}{ c | c | c | c | c }
		\hline\noalign{\smallskip}
		${7\over 2}[633]$ & & & & \\
		
		$|N l j \Omega \rangle$ & 
		$\left| 6 4 {7\over 2} {7\over 2} \right\rangle$ & 
		$\left| 6 4 {9\over 2} {7\over 2} \right\rangle$ & 
		$\left| 6 6 {11\over 2} {7\over 2} \right\rangle$ &
		$\left| 6 6 {13\over 2} {7\over 2} \right\rangle$ \\
		
		$\epsilon$ & & &  \\
		
		\noalign{\smallskip}\hline\noalign{\smallskip}
		
		0.05 & $-0.0012$ & 0.0544 & $-0.0166$ & 0.9984  \\
		0.22 & $-0.0161$ & 0.1927 & $-0.0744$ & 0.9783  \\
		0.30 & $-0.0254$ & 0.2380 & $-0.1000$ & 0.9658  \\
		
		\noalign{\smallskip}\hline
	\end{tabular}
	
	\begin{tabular}{ c  c | c | c | c }
		\hline\noalign{\smallskip}
		${9\over 2}[624]$ & & & & \\
		
		$|N l j \Omega \rangle$ & 
		& 
		$\left| 6 4 {9\over 2} {9\over 2} \right\rangle$ & 
		$\left| 6 6 {11\over 2} {9\over 2} \right\rangle$ &
		$\left| 6 6 {13\over 2} {9\over 2} \right\rangle$ \\
		
		$\epsilon$ & & &  \\
		
		\noalign{\smallskip}\hline\noalign{\smallskip}
		
		0.05 &  & 0.0336 & $-0.0176$ & 0.9993  \\
		0.22 &  & 0.1120 & $-0.0724$ & 0.9911  \\
		0.30 &  & 0.1366 & $-0.0947$ & 0.9861  \\
		
		\noalign{\smallskip}\hline
	\end{tabular}
	
	\begin{tabular}{ r r r r r }
		\hline\noalign{\smallskip}
		${5\over 2}[512]$ & & & & \\
		
		$|N l j \Omega \rangle$ & 
		$\left| 5 3 {5\over 2} {5\over 2} \right\rangle$ & 
		$\left| 5 3 {7\over 2} {5\over 2} \right\rangle$ & 
		$\left| 5 5 {9\over 2} {5\over 2} \right\rangle$ &
		$\left| 5 5 {11\over 2} {5\over 2} \right\rangle$ \\
		
		$\epsilon$ & & &  \\
		
		\noalign{\smallskip}\hline\noalign{\smallskip}
		
		0.05 &    0.0659 & 0.2016 & 0.9772 &    0.0047  \\
		0.22 & $-0.0242$ & 0.8371 & 0.5231 & $-0.1580$  \\
		0.30 & $-0.0619$ & 0.8684 & 0.4439 & $-0.2123$  \\
		
		\noalign{\smallskip}\hline
	\end{tabular}
	
\end{table}

\begin{table}[htb]
\centering
\caption{Expansions of proton Nilsson orbitals $\Omega[N n_z \Lambda]$ in the shell model basis $|N l j \Omega \rangle$ for three different values of the deformation $\epsilon$. 
	}
	\label{tab:4}       
	
	\begin{tabular}{ r r r r  }
		\hline\noalign{\smallskip}
		${7\over 2}[523]$ & & & \\
		
		$|N l j \Omega \rangle$ & 
		$\left| 5 3 {7\over 2} {7\over 2} \right\rangle$ & 
		$\left| 5 5 {9\over 2} {7\over 2} \right\rangle$ & 
		$\left| 5 5 {11\over 2} {7\over 2} \right\rangle$ \\

		$\epsilon$ & & &  \\
		
		\noalign{\smallskip}\hline\noalign{\smallskip}
		
		0.05 & 0.0323 & $-0.0212$ & 0.9993  \\
		0.22 & 0.1129 & $-0.0872$ & 0.9898  \\
		0.30 & 0.1398 & $-0.1138$ & 0.9836  \\
		
		\noalign{\smallskip}\hline
	\end{tabular}
	
	\begin{tabular}{ r r r  }
		\hline\noalign{\smallskip}
		${9\over 2}[514]$ & &  \\
		
		$|N l j \Omega \rangle$ &                                                 
		$\left| 5 5 {9\over 2} {9\over 2} \right\rangle$ & 
		$\left| 5 5 {11\over 2} {9\over 2} \right\rangle$ \\
		
		$\epsilon$ & &  \\
		
		\noalign{\smallskip}\hline\noalign{\smallskip}
		
		0.05 & $-0.0194$ & 0.9998  \\
		0.22 & $-0.0716$ & 0.9974  \\
		0.30 & $-0.0907$ & 0.9959  \\
		
		\noalign{\smallskip}\hline
	\end{tabular}
	
\end{table}

\begin{acknowledgments}
 The author J. D. V is  indebted to Rick  Casten for useful comments and suggestions.
 The authors thank Mr. Gannon Lawley for his helpful context cross check. 
\end{acknowledgments}

\bibliography{main}

\end{document}